\documentclass[a4paper,11pt]{article}
\pdfoutput=1 

\usepackage{longtable,booktabs,caption,amsmath}

\usepackage[skip=10pt plus1pt, indent=5pt]{parskip}
\usepackage{jcappub} 
\usepackage[T1]{fontenc}

\usepackage{parskip}
\usepackage{mathtools}
\usepackage{enumitem}
\usepackage{dsfont}
\usepackage{xcolor}
\definecolor{greyforboxes}{RGB}{235,235,235}

\usepackage{tikz}                           
\usepackage[framemethod=tikz]{mdframed}      
\mdfsetup{splittopskip=\topskip, linewidth=0.5}

\usepackage{amsmath}

\def\be{\begin{equation}}
\def\ee{\end{equation}}
\def\bea{\begin{eqnarray}}
\def\eea{\end{eqnarray}} 

\definecolor{darkyellow}{rgb}{0.5, 0.5, 0.0}
\definecolor{darkpurple}{rgb}{0.5, 0.2, 0.8}
\definecolor{darkblue}{rgb}{0.0, 0.0, 0.8}
\definecolor{darkgreen}{rgb}{0.0, 0.4, 0.0}
\definecolor{darkred}{rgb}{0.5, 0.0, 0.0}
\usepackage{hyperref}
\hypersetup{
    linktocpage,
     colorlinks,
     citecolor=darkgreen,
     linkcolor= darkgreen,
     urlcolor=darkgreen
}

 \notoc 

\begin{document}
\hskip12.5cm  ET-0192A-22
\title{\huge On the response of  the Einstein Telescope\\ to Doppler  anisotropies}
\author[a]
{Debika Chowdhury,}
\author[a,b]
{Gianmassimo Tasinato,}
\author[a]
{Ivonne Zavala}
\affiliation[a]
{Department of Physics, Swansea University, Swansea, SA2 8PP, United Kingdom}
\affiliation[b]
{Dipartimento di Fisica e Astronomia, Universit\`a di Bologna, via Irnerio 46, Bologna,  Italy}
\emailAdd{debika.chowdhury at swansea.ac.uk} 
\emailAdd{g.tasinato2208 at gmail.com}
\emailAdd{e.i.zavalacarrasco at swansea.ac.uk}

\abstract{
We study the response function of the Einstein Telescope
to kinematic Doppler anisotropies, which represent one of the 
guaranteed properties of the stochastic gravitational wave background.
If the frequency dependence 
of the stochastic background changes slope within the detector frequency
band, the Doppler anisotropic
contribution to the signal 
  can   not be factorized in a part 
depending on frequency, and a part depending on direction. 
 For the first time, we study the detector response
function  to Doppler anisotropies without making any factorizable Ansatz. 
Moreover, we do not assume that kinematic effects are small, and we derive
general formulas   valid for any relative velocity among frames. We 
apply our findings to three well-motivated examples of background profiles:
power-law, broken power-law, and models with a resonance motivated
by primordial black hole scenarios. We  derive the 
signal-to-noise
ratio associated with  an optimal estimator for the detection of 
non-factorizable
kinematic anisotropies,
 and we study it for
 representative examples. 
  }

\maketitle

\section{\color{black} Introduction }

Once a stochastic gravitational wave background (SGWB) will be detected -- see \cite{Renzini:2022alw} for current
prospects --
the next goal for GW science will be to characterize its anisotropies.  As 
for the cosmic microwave background (CMB), the anisotropies of the SGWB
promise  to provide  information on the origin and evolution
of the  GW signal. SGWB anisotropies can be produced by the 
 mechanisms that source the SGWB 
\cite{Olmez:2011cg,Kuroyanagi:2016ugi,Cusin:2017fwz,Jenkins:2018nty,Cusin:2018rsq,Jenkins:2018uac,Cai:2021dgx,Geller:2018mwu,Pitrou:2019rjz,Capurri:2021zli,Dimastrogiovanni:2021mfs,Dimastrogiovanni:2022afr,Dimastrogiovanni:2022eir}, 
 or by propagation effects  through a perturbed
universe \cite{Alba:2015cms,Contaldi:2016koz,Bartolo:2019oiq,Bartolo:2019yeu,Bertacca:2019fnt,Bartolo:2019zvb,ValbusaDallArmi:2020ifo,Domcke:2020xmn}.  Alternatively, they can have a  kinematical
origin, being induced by  the detector motion with velocity $\vec v$
with respect to the rest frame of the SGWB. In this work
we focus on this last kind   of  SGWB anisotropies, which have been recently theoretically
 investigated 
  in \cite{LISACosmologyWorkingGroup:2022kbp,Cusin:2022cbb,LISACosmologyWorkingGroup:2022jok} (see also \cite{DallArmi:2022wnq,Chung:2022xhv} for applications and further developments).

The fact that Doppler anisotropies can be  relevant for observations of  stochastic backgrounds 
is made manifest by the CMB kinematic
dipole,  whose amplitude is around two orders of magnitude
larger than that of CMB intrinsic anisotropies \cite{Smoot:1977bs,Kogut:1993ag,WMAP:2003ivt,Planck:2013kqc}. For the case of SGWB, in 
absence of detection, we do not yet know how sizeable SGWB Doppler anisotropies can be.
The  velocity $\vec v$ among the SGWB and our frames could  be large: think for example of
a  SGWB produced in the  early universe, during a phase transition within  
 a cosmic fluid in relativistic 
  coherent motion (see e.g. \cite{Caprini:2018mtu,Regimbau:2011rp} for general reviews on SGWB sources). 
  
At the moment,  given our ignorance on possible sources of SGWB, it is then wise to keep non-committal on the 
 relative speed $\vec v$,  and on the SGWB intrinsic properties.   
   In order to forecast prospects
  of detection of Doppler anisotropies, the first step is to investigate the  response
  of GW experiments to their possible  features.  
Previous articles studied in  detail the response of GW detectors
to  anisotropies, starting with  \cite{Allen:1996gp,Cornish:2001hg,Ungarelli:2001xu,Seto:2004np,Kudoh:2004he,Taruya:2005yf,Taruya:2006kqa,Thrane:2009fp,Mingarelli:2013dsa,Taylor:2013esa,Baker:2019ync} (see \cite{Romano:2016dpx} for a general review).  Usually, one assumes    a {\it factorizable Ansatz} for the
quantities describing the anisotropic signal. The signal should be   described in terms of  a contribution depending
on GW frequency, times a contribution  depending on GW direction only. However, in general,  such an Ansatz
is not suitable for describing Doppler anisotropies. In fact, building on \cite{Cusin:2022cbb}, we show
explicitly that if the SGWB slope changes within the detector frequency band -- a very common
possibility both for astrophysical and cosmological sources (see e.g. \cite{Bartolo:2016ami} in the context of LISA) --  the aforementioned
factorizable Ansatz is violated. 

In this work we outline
a method for studying the general case, making use of special simplifying 
 conditions  (first pointed out
 in \cite{Mentasti:2020yyd}) 
 for 
 characterizing the study of SGWB with the Einstein Telescope \cite{Punturo:2010zz,Maggiore:2019uih}: 
 we apply it to the case of kinematic anisotropies. Our method does not  implement
 any factorizable Ansatz, nor makes the hypothesis that the speed $\vec v$ among
 frames is small. 
  We are able to 
  express the detector response function to an anisotropic
 signal in terms of   combinations  of detector properties (the arm directions)
 contracted with the velocity vector $\vec v$: see sections \ref{sec_setup} and \ref{sec_kin}.   In particular, we find 
 that the Einstein Telescope
response depends in a non-linear (but  computable)
  way on  $\vec v$ as well as on the frequency dependence of the SGWB profile.      
  As a byproduct,
   our findings
  indicate that   SGWB Doppler anisotropies can provide us with independent
   measurements of key features of the SGWB.

We proceed   in
section
 \ref{sec_threex} investigating 
the  
 ET response function to three explicit  examples of SGWB with well motivated frequency profiles: power
 law, broken power law (see e.g. \cite{Kuroyanagi:2018csn} for a survey), and models with a resonance,  motivated by second-order SGWB induced by the formation of primordial black holes  (see e.g. \cite{Ananda:2006af,Baumann:2007zm,Saito:2008jc,Saito:2009jt}). In section
 \ref{sec_snr} we determine an optimal estimator for detecting Doppler anisotropies,  exploiting
 the characteristic daily modulation   of kinematic anisotropies \cite{Allen:1996gp}. We 
 obtain  the expression for the corresponding optimal signal-to-noise
 ratio, and we apply our results to  representative examples of broken power law SGWB profiles, in order
 to investigate explicitly how the frequency dependence 
of the SGWB affects measurements of Doppler effects. Our conclusions in section \ref{sec_conc}
 discuss  possible   further developments, and are followed by four technical appendixes.

\section{\color{black} Our setup}
\label{sec_setup}

In order to characterise  the interferometer
response to Doppler anisotropies, we first 
present some basic ingredients, 
and fix the conventions we use for describing 
 the stochastic gravitational
wave background (SGWB). 
We expand  the gravitational wave perturbation  $h_{ab}(t, \vec x)$ in Fourier modes as (setting $c=1$)
\be
h_{ab}(t, \vec x)\,=\,\sum_\lambda 
\int_{-\infty}^{+\infty}\,d f\,\int d^2 \hat n\,h_\lambda(f, \hat
n)\,{\bf e}^{(\lambda)}_{ab}(\hat n)
\,e^{2 \pi\,i\,f (t-\hat n \cdot \vec x)}\,,
\ee
where $\lambda\,=\,(+, \,\times)$ are the polarization indices, 
$f$ the GW frequency, and $\hat n$ a unit vector indicating the GW direction. To ensure that the GW fluctuation is real, we impose $h_\lambda(f, \hat
n)\,=\,h^*_\lambda(-f, \hat
n)$.  The polarization tensors satisfy the condition ${\bf e}^{(\lambda)}_{ab}(\hat n)\,=\,{\bf e}^{(\lambda)\,\,*}_{ab}(-\hat n)$, and are normalized such that ${\bf e}^{(\lambda)}_{ab}(\hat n)\,{\bf e}^{(\lambda')}_{ab}(\hat n)\,=\,2\, \delta^{\lambda \lambda'}$. The two-point
correlator for the Fourier
modes $h_\lambda(f, \hat n)$ reads
\be
\label{def_2pt}
\langle h^*_{\lambda}(f,\hat n)\, h_{\lambda'}(f',\hat n')
\rangle\,=\,\frac{\delta_{\lambda \lambda'}}{2}\,\delta(f-f')\,\frac{\delta^{(2)}(\hat n-\hat n')}{4 \pi}\,{\cal I}(f)\,{\bf P}(f, \hat n)\,,
\ee
with ${\cal I}(f)$ being the GW intensity,    an even function of frequency: ${\cal I}(f)\,=\,{\cal I}(|f|)$. The
factor ${\bf P}(f, \hat n)$ 
  accounts  for the SGWB  anisotropy. Importantly,  {\it we do not assume a factorizable Ansatz} for ${\bf P}$, and we 
  allow it to be  an arbitrary function of $f$ and $\hat n$. Our  general treatment will suit our analysis of
   kinematic anisotropies in section \ref{sec_kin}. Nevertheless, we assume 
  for simplicity that 
  ${\bf P}=1$ if the two-point correlator \eqref{def_2pt} is isotropic, i.e. when the function ${\bf P}$
  is independent of the GW direction $\hat n$.

\medskip

The quantity we are interested in is the 
laser phase difference  as measured
by a planar ground-based 
 interferometer: in particular, we have in mind
the Einstein Telescope (ET) \cite{Punturo:2010zz}. To 
characterize this quantity, we 
  follow the discussion of   \cite{Smith:2019wny} developed for LISA: we adapt 
it to the case of an anisotropic GW background measured
by a ground-based instrument. We consider two arms $AB$ and $AC$ of the interferometer,
and  indicate  with $\Phi_{A_{BC}}$ the phase-difference,  as measured 
 at the common vertex $A$. Such a phase-difference can be decomposed into two parts: 
 \be
\Phi_{A_{BC}}(t)\,=\,\Delta \varphi_{A_{BC}}(t)+n_{A_{BC}}(t)\,.
\ee 
In this expression, $\Delta \varphi_{A_{BC}}$ is the GW contribution (if any), and $n_{A_{BC}}$ is the noise. 
Introducing  a standard nomenclature for GW physics (see e.g. \cite{Maggiore:2007ulw}),  
the GW
contribution can be expressed as 
\be
\Delta \varphi_{A_{BC}}(t)\,=\,\sum_\lambda\,\int_{-\infty}^{\infty}
d f\,\int d^2 \hat n\, h_{\lambda}(f,\hat n)\,e^{2 \pi i\,f t}\,F_{A_{BC}}^{(\lambda)}
(t, \hat n)\,,
\ee
where $F_{A_{BC}}^{(\lambda)}$ is the detector pattern function. We work in a small-frequency limit, suitable
for ET \cite{Mentasti:2020yyd} for which $2 \pi f\,L\ll1$, with $L$ being the detector arm length. The detector
pattern function reads 
\be
F_{A_{BC}}^{(\lambda)}(\hat n, t)\,=\,e^{-2 \pi\,i\,f\,\hat n \cdot \vec x_A(t)}
{\bf e}_{ab}^{(\lambda)}(\hat n)\,
d_{A_{BC}}^{ab}(t)\,,
\ee
with $d_{A_{BC}}^{ab}$ being the detector tensor, 
\be
\label{def_dAB}
d_{A_{BC}}^{ab}(t)
\,=\, \frac12\,\left(\ell_{AB}^a(t)
 \ell_{AB}^b(t)-
 \ell_{AC}^a(t)
 \ell_{AC}^b(t)
 \right)\,,
\ee
and  $ \ell_{AB}(t)$ indicating   the unit vector pointing between $A$ and
$B$ vertexes. Notice that the quantity $d_{A_{BC}}^{ab}$ is traceless:  
$d_{A_{BC}}^{aa}\,=\,0$.

 Assuming that noise and GW signals are uncorrelated,
the two-point correlation  function among phase differences can be expressed as
\be
\label{def_2ptlas}
\langle \Phi_{A_{BC}}(t)
\Phi_{X_{YZ}}(t')
\rangle
\,=\,\frac12\,\int_{-\infty}^{\infty}
\,d f\,e^{2 \pi i\,f (t-t')}
\,\left[ {\cal R}_ {A_{BC}, X_{YZ}}(f, t, t')\,{\cal I}(f)+   N_{A_{BC}, X_{YZ}}(f)\right]
\,.
\ee

Here $X_{YZ}$ denotes a vertex $X$ between two interferometer arms $XY$ and $XZ$. Those arms can
belong to the same instrument as the arms $AB$ and $AC$ (i.e. a single version of
the  ET interferometer),
or instead to a second independent ET-like instrument, as discussed in \cite{Mentasti:2020yyd}, in order
to reduce the correlated noise. Our arguments can in principle 
apply to both situations.  In eq \eqref{def_2ptlas}, ${\cal I}$ is  the GW intensity as introduced in eq \eqref{def_2pt},
while 
$ N_{A_{BC}, X_{YZ}}$ the variance of the noise Fourier transform:
\be
\label{def_noise}
\langle
\tilde n_{A_{BC}}^* \,\tilde n_{X_{YZ}}
\rangle\,=\,
\delta(f-f')\,\delta^{(2)}(\hat n-\hat n' )
\,
N_{A_{BC}, X_{YZ}}\,.
\ee
The function
  $ {\cal R}_ {A_{BC}, X_{YZ}}$ in eq \eqref{def_2ptlas} is
the detector response function that we wish to characterize.  Collecting
the results,  such a response function can be expressed as
\bea
\label{def_res1}
 {\cal R}_ {A_{BC}, X_{YZ}}(f, t, t')
 &=&\sum_\lambda\,\int \frac{d^2 \hat n}{4 \pi}
 \,{\bf P}(f, \hat n)\,
  F_{A_{BC}}^{(\lambda)}(\hat n, t)
  F_{X_{YZ}}^{(\lambda)}(\hat n, t')
 \nonumber
 \\
 &=&
 {d^{ab}_{A_{BC}}(t)}
 \,
{d^{cd}_{X_{YZ}}(t')}
\,
\Gamma_{abcd}(f)
\label{def_res1a}\,,
\eea
with $d^{ab}$ given in eq \eqref{def_dAB}, while
\be
\label{def_gamma}
\Gamma_{abcd}(f)\,=\,\sum_{\lambda}
\,\int \frac{d^2 \hat n}{4 \pi}\,e^{-2 \pi\,i\,f \,\hat n \cdot \Delta \vec x}{\bf P}(f,\,\hat n)\,{\bf e}_{ab}^{(\lambda)}(\hat n)
{\bf e}_{cd}^{(\lambda)}(\hat n)\,.
\ee
In eq \eqref{def_gamma}, 
 $ \Delta \vec x$ denotes the spatial difference between the vertexes $A$ and $X$.
The response function as defined above depends on the direction-dependent quantity ${\bf P}$ as introduced in \eqref{def_2pt}, and  controls the anisotropy of the GW correlator. The quantity $\Gamma_{abcd}(f)$ is symmetric under the interchanges $a \leftrightarrow b$,  $c \leftrightarrow d$,  $a b \leftrightarrow c d$. Moreover, $\Gamma_{aacd}(f)=0$. 
Eq \eqref{def_res1} is an extension of  well-known formulas (see e.g. \cite{Allen:1997ad}) to 
the case of anisotropic SGWB.

The covariant matrix of phase differences in each vertex can be diagonalized as explained in \cite{Mentasti:2020yyd}
in the context of the ET interferometer.
 We refer the reader to this work for
 details; we do not have anything to add to this topic.   After diagonalization, one determines three diagonal
channels (called  $A$, $E$, $T$),   denoted with the letters ${\cal O}, {\cal O}'$.

 We  aim at characterizing
the response function ${\cal R}_ {{\cal O}, {\cal O}' }(f, t)$ for each diagonal
channel: for doing so, we need to analyze the structure
of the quantity $\Gamma_{abcd}$ of eq \eqref{def_gamma}.  For the case of the Einstein Telescope,   a major simplification arises, as first found and exploited in  \cite{Mentasti:2020yyd}. Since the instrument is mostly sensitive to relatively small values of the frequency, $f \sim 7$ Hz, 
 the exponent
depending on the vertex distance in eq \eqref{def_gamma} can be neglected. 
 Indeed, 
 we have 
$
|2 \pi\,f \,\Delta \vec x |\, \simeq\,6\times 10^{-5}\,\left(\frac{f}{\rm Hz}\right)
\left( \frac{\Delta x}{\rm km}\right)\,
$.
This quantity is small (at most of order of percent) for correlations between the  arms of a single ET interferometer; or, for
%
%
 correlations between two distinct interferometers
located at different places on the Earth surface (but say within continental Europe, see \cite{Mentasti:2020yyd}).  Under this approximation, in what comes next we are
going to determine the structure of the response function of the ET interferometer to Doppler anisotropies of SGWB, with  
no need to make any factorizable Ansatz for ${\bf P}(f, \hat n)$, or to
assume that kinematic effects are perturbatively small.  In fact,
we  elaborate a method 
allowing  us to compute  \eqref{def_gamma}, with no need of any expansion 
in spherical harmonics (which is not too  well suited for general scenarios where ${\bf P}(f,\,\hat n)$ is explicitly
frequency-dependent, as ours).

\section{\color{black} ET response function to kinematic anisotropies}
\label{sec_kin}

Kinematic anisotropies arise from the motion of our GW detector
with respect to the rest frame of the SGWB source. These Doppler
effects are expected
to occur for any background of primordial or astrophysical origin, and
can provide the largest anisotropic contribution to a SGWB signal. As a concrete
example, for
the CMB the amplitude of the kinematic dipole is two
orders of magnitude larger than the typical size of intrinsic
CMB anisotropies of primordial origin. 

For the case of the SGWB, since we are ignorant about its source (if any)
and its velocity with respect to us,
we prefer to keep  non-committal,  and derive general formulas which can be
applied to generic  situations we might encounter. Data, if and when available,  
will provide information about the relative velocity among frames. 
We make use
of the  analytic formulas for kinematic anisotropies
recently derived in \cite{Cusin:2022cbb}, to obtain results that are
valid for any speed of the GW source with respect to us, and for
any frequency profile of SGWB signal. (We make the simplifying
hypothesis, though, that the GW signal is isotropic in the
rest-frame of the SGWB source.)

  \medskip
  We indicate with 
 $\Omega_{\rm GW}^{(A)}(f)$ the GW
 energy density in the rest frame of the SGWB source: as mentioned above, we assume
 it to be isotropic. The GW energy density becomes  anisotropic in a boosted frame $(B)$ moving with velocity $\vec v$
 wrt $(A)$. In fact,  denoting with
 $\beta\,=\,|\vec v|$ (in units with $c=1$) the size of the relative velocity among frames,  
$ \Omega_{\rm GW}^{(B)}$ results  \cite{Cusin:2022cbb}
 \be
 \label{anOGW}
 \Omega_{\rm GW}^{(B)}(f, \hat n)\,=\,{\cal D}^{4}\,
 \Omega_{\rm GW}^{(A)}\left({\cal D}^{-1}
 \,
 f\right)\,,
  \ee
  with
  \be
  \label{def_D}
  {\cal D}\,=\,\frac{\sqrt{1-\beta^2}}{1-\beta\, \hat n \cdot \hat v}\,,
  \ee
  and $\hat n$ and $\hat v$ are the unit
  vectors along GW direction and the relative velocity of the frame, respectively.  Hence, $ \Omega_{\rm GW}^{(B)}$ in
  eq \eqref{anOGW} is anisotropic, and in general the effects of anisotropy
{\it   can not} be factorized in a part depending on frequency,  and another one on direction \cite{Cusin:2022cbb}.
Hence 
  the analysis as commonly carried on in   previous works should be accommodated to the present situation. 
  Recall that, for the isotropic case, the SGWB energy dependence is related to the GW intensity ${\cal I}(f)$
  by 
  \be
  \label{def_Ois}
   \Omega^{\rm isotropic}_{\rm GW}(f)\,=\,\frac{ 4 \pi^2\,f^3}{3\,H_0^2}\,{{\cal I}(f)}\,.
   \ee
  with $H_0$ being the present-day Hubble parameter. 
   Using eq \eqref{anOGW}, we can then conveniently 
  express the GW energy densities in frames $(A)$ and $(B)$ as
 \be
 \Omega_{\rm GW}^{(A)}(f)\,=\,\frac{ 4 \pi^2}{3\,H_0^2}\,f^3\,{\cal I}(f)
 \hskip0.6cm; \hskip0.6cm
 \Omega_{\rm GW}^{(B)}(f)\,=\,\frac{ 4\pi^2}{3\,H_0^2}\,f^3\,{\cal I}(f)
 \,{\bf P}_{\rm kin}(f,\,\hat n)\,,
 \ee
 with (recall the definition \eqref{def_D})
 \bea
 \label{res_P}
 {\bf P}_{\rm kin}(f,\,\hat n)&=&\frac{{\cal D}\,}{ {{ \cal I}}(f)}
 \,{{\cal I}}({\cal D}^{-1}\,f)
 \label{def_bfP}\,.
 \eea
The expression  in eq \eqref{res_P} demonstrates explicitly that    ${\bf P}_{\rm kin}(f,\,\hat n)$  can not be factorized
 in a part depending on frequency, times a part  depending on direction (unless  ${{\cal I}}$ is an exact
 power law). 
 In fact, if the frequency profile of ${\cal I}$ changes within the detector frequency band, the
 dependence  of  \eqref{res_P}  on ${\cal D}$ (hence on the anisotropy vector $\hat v$) changes at the positions
 where the slope of  ${\cal I}$ changes. We will discuss examples of this possibility in the next sections.
  Notice   that if  $\beta=0$ (no kinematic anisotropy) then ${\bf P}_{\rm kin}\,=\,1$,  as desired.
  
  \medskip
  Fortunately, given the special properties of ET \cite{Mentasti:2020yyd}
  which we discussed at the end of section \ref{sec_setup}, we can nevertheless derive an exact expression for the ET response
  function to kinematic anisotropies, with no need of simplifying Ans\"atze.  
  As we show in the technical appendix \ref{app_A}, the interferometer response function
   \eqref{def_res1a} relative to the $A, E, T$ channels can be expressed as a combination of three terms only,  with transparent geometrical meanings:
   \bea
  {\cal R}_ {{\cal O},\,{\cal O}'}(f, t, t')
 &=&\frac45\left[1 +\frac{5}{2} c_1(f) \right]\,{d^{ab}_ {\cal O}(t)
 \,
d_ {{\cal O'}\,ab}(t')}
\nonumber
\\
&&
+4\,c_2(f)\,\left(
{\hat v}_a\,{d^{ab}_{{\cal O}}(t)
 \,
d_{{\cal O}'\,bc}(t')}\,{\hat v}^c
\right)
\nonumber
\\
&&+c_3(f)\,\left(\hat v^a \hat v^b\,d_{ab\,{\cal O}}(t)\right)\,\left(\hat v^c \hat v^d\,d_{cd\,{\cal O'}}(t')\right)\,,
\label{res_OOp}
 \eea

\noindent
with ${\cal O},\,{\cal O}'$ denoting  the interferometer channels,  $d^{ab}$ being the detector tensor \eqref{def_dAB}, and $\hat v$ 
the unit velocity vector  among the two frames.  
The quantities introduced in \eqref{res_OOp} read

\bea
 c_1&=&\frac{K_1}{8}+\frac{3\,K_2}{4}+\frac{K_3}{8}\hskip0.2cm;\hskip0.2cm
    c_2\,=\,\frac{3\,K_1}{8}-\frac{3\,K_2}{4}-\frac{5\,K_3}{8}
   \hskip0.2cm;\hskip0.2cm
  c_3\,=\,\frac{3\,K_1}{8}-\frac{15\,K_2}{4}+\frac{35\,K_3}{8}\,,
  \nonumber\\
  \label{def_ci}
 \eea
 where
 \bea
 K_1&=&\int \frac{d^2 \hat n}{4 \pi} \left( {\bf P}_{\rm kin}-1\right)\hskip0.3cm;\hskip0.3cm
  K_2 = \int \frac{d^2 \hat n}{4 \pi} \left( {\bf P}_{\rm kin}-1\right)\,(\hat n \cdot \hat v)^2
  \hskip0.3cm;\hskip0.3cm
  K_3= \int  \frac{d^2 \hat n}{4 \pi} \left( {\bf P}_{\rm kin}-1\right)\,(\hat n \cdot \hat v)^4\,.
\nonumber\\
\label{def_intK}
 \eea
 Some comments on the results so far:
 \begin{itemize}
 \item All the effects
 of kinematic anisotropies in the response function \eqref{res_OOp} are contained in the
 three terms proportional to the frequency-dependent
 coefficients $c_{i}(f)$ in eq \eqref{def_ci}. They
 depend on covariant contractions of the detector tensors
 $d^{ab}(t)$ with the direction $\hat v^a$ of the relative frame velocity.
 Their frequency dependence has important implications when
 discussing perspectives of detection, as we will discuss in what follows.
 \item
 The
 three  independent terms of the response function  \eqref{res_OOp} resemble in spirit the effects of the  three multipoles
  $\ell\,=\,(0,\,2,\,4)$ that were found
  in \cite{Mentasti:2020yyd} to contribute to anisotropies
   detectable by ET.
   We refrain from elaborating
   on this analogy in this work, since in our approach we do not implement a multipolar expansion
   of the anisotropic signal, given that  we can not factorize it in frequency times direction.  
   Nevertheless, it would 
   be  interesting to understand whether some alternative  generalization of the approach of \cite{Mentasti:2020yyd} to a non-factorizable Ansatz can lead to  results as  ours. 
   %
%
   %
  \item Our method relies  on the 
  computation of the three integrals $K_i$
  in eq \eqref{def_intK}, which can  
  be easily performed by  numerical tools, depending on the
  profile of ${\cal I}(f)$ (recall the definition of $ {\bf P}_{\rm kin}$ in eq \eqref{res_P}). 
  Notice that they all vanish when $\beta\,=\,0$ (no
  kinematic effects) or when ${\cal I}(f)$ is
  a linear function of frequency (see eq \eqref{def_bfP}) \footnote{In
  fact, then $\Omega_{\rm GW}$
  is proportional to $f^4$, 
  a particular case in which kinematic
  effects cancel out \cite{Cusin:2022cbb}.}.
  But in general, our formulas are
  valid for any size of $0\le \beta<1$,
  and encompass all kinematic effects 
  with no need of any perturbative expansion in $\beta$.  
  \item The geometrical
   quantities appearing in the response function \eqref{res_OOp}
   explicitly depend on time: in particular, the orientation
   of the detector(s) with respect to the velocity vector $\hat v$
   experiences daily and annual modulations due to the motion
   of the Earth. This property
   will be crucial for determining the optimal estimator sensitive to 
   kinematic anisotropies \cite{Allen:1996gp}: see  section \ref{sec_snr}. Interestingly,
   our general formulas can also describe scenarios
   where the kinematic
   parameters $\beta$ and $\hat v$ have intrinsic time-dependence
   (not just due  to the  Earth motion).   
   It would be interesting in future works to explore whether there can be
   SGWB sources realizing this possibility.
   \end{itemize}

  \noindent
 In the next section \ref{sec_threex} we study  three well-motivated
 examples 
 of ${\cal I}(f)$, so as to concretely explore the effects of kinematic
 anisotropies on the response function of  ET.

   \section{\color{black}Three examples of SGWB frequency profiles} 
   \label{sec_threex}

We  apply our methods to three
well motivated scenarios for the  frequency dependence of  ${\cal I}(f)$. We start discussing
the case of an exact power-law profile, for which we are able to obtain fully analytic
formulas valid for any value of $0\le\beta<1$. We then continue discussing the cases
of single and multiple broken power law, for which the function
   ${\bf P}_{\rm kin}$ --
the quantity controlling the SGWB anisotropy --  is not  factorizable in 
  parts depending respectively on frequency and   direction. 
In such cases, we compute how anisotropies depend
on  the frequency profiles of  ${\cal I}(f)$, with no restrictions on the size of $\beta$ within the interval $0\le\beta<1$.

 \subsection{\color{black}First example: a  power-law SGWB profile} 
 \label{sec_firex}
 
 We start by considering a power-law  intensity profile in the SGWB rest frame,
 as described by the Ansatz
 \be
 \label{ans_pl1}
 {\cal I}^{PL}(f)\,=\,I_0 \,\left(\frac{f}{f_\star}\right)^\alpha\,,
 \ee
where $I_0 $ is a normalization factor, and $f_\star$ is a reference frequency.
 Relation \eqref{ans_pl1} for the intensity implies, through eq \eqref{def_Ois}, that $\Omega_{\rm GW}$ scales with frequency as
 \be
 \label{scalOM}
 \Omega_{\rm GW}\,\propto\,f^{3+\alpha}\,.
 \ee

The degree of kinematic anisotropy depends on the parameter $\alpha$
in eq \eqref{ans_pl1}.
  Making use of eq \eqref{def_bfP},  the kinematic anisotropy parameter  ${\bf P}_{\rm kin}$ reads
   \be
 {\bf P}_{\rm kin}\,=\,
 \frac{ {\cal D} }{
   {{ \cal I}^{PL}}(f)}\,
\,{{\cal I}^{PL}}({\cal D}^{-1}\,f)\,=\, 
{\cal D}^{1-\alpha},
\ee
confirming that, in this
particular  case, the dependence on frequency cancels out. The integrals \eqref{def_intK} can be done analytically, and we
can build
exact expressions for the quantities $c_{1,\,2,\,3}$ which enter in the response function
of eq \eqref{res_OOp}, for any values of $\alpha$ and $0\le\beta\le1$. The
complete formulas are rather long and we relegate them to Appendix \ref{app_B}. 
  In table \ref{Table3} we present the exact results for the coefficients $c_i$ given in eq \eqref{def_ci} 
  as functions of $\beta$. We make  three representative choices of the exponent $\alpha$.
  
  \begin{table}[h!]
\begin{center}
\centering
\begin{tabular}{| l | c | c | c|}
\hline
&  $\alpha=-3$ & $\alpha=3$ & $\alpha=5$   \\
\hline 
\hline
$c_1$ &    $-\frac{64 \beta ^7+66 \beta ^5-80 \beta ^3+30 \left(\beta ^2-1\right)^3 \tanh ^{-1}(\beta )+30 \beta }{60 \beta ^5 \left(\beta ^2-1\right)}$ 
&  $\frac{64}{105}\,\frac{\beta^2}{1-\beta^2}$  &
$\frac{8\,\beta^2}{315}\,\frac{(81-10\,\beta^2)}{\left(1-\beta^2\right)^2}$
\\ 
\hline
$c_2$ & $\frac{66 \beta ^5-80 \beta ^3+30 \left(\beta ^2-1\right)^3 \tanh ^{-1}(\beta )+30 \beta }{12 \beta ^5 \left(\beta ^2-1\right)}$
&
$-\frac{4}{35}\,\frac{\beta^2}{1-\beta^2}$
&
$-\frac{8\,\beta^2}{315}\,\frac{(27+4\,\beta^2)}{\left(1-\beta^2\right)^2}$ 
\\ 
\hline
$c_3$ &
$\frac{96 \beta ^7-462 \beta ^5+560 \beta ^3-210 \left(\beta ^2-1\right)^3 \tanh ^{-1}(\beta )-210 \beta }{12 \beta ^5 \left(\beta ^2-1\right)}$
& $0$ 
& $\frac{8}{315}\,\frac{\beta^4}{\left(1-\beta^2\right)^2}$
\\
\hline
\end{tabular}
\end{center} 
\caption{The   $c_i$ of eq \eqref{def_ci} for three choices 
of exponents in the power-law Ansatz of eq \eqref{ans_pl1}.}
\label{Table3}
\end{table}   
In each case, the absolute value of the size of the anisotropy contributions to the detector response
function  monotonically increases,  as $\beta$ increases towards $\beta\to1$. 
The case $\alpha=-3$ corresponds to a scale-invariant GW density parameter,
according to eq \eqref{scalOM}.

Notice that,
in the small
$\beta$ limit, contributions start only at order $\beta^2$ (or higher): the ET response
is insensitive to linear contributions in $\beta$ to kinematic anisotropies, corresponding
to the kinematic dipole. 
This reflects the fact that ET is insensitive to the dipolar anisotropies \cite{Mentasti:2020yyd},
at least within the approximation we are interested in.

\begin{figure}[h!]
\centering
  \includegraphics[width = 0.48 \textwidth]{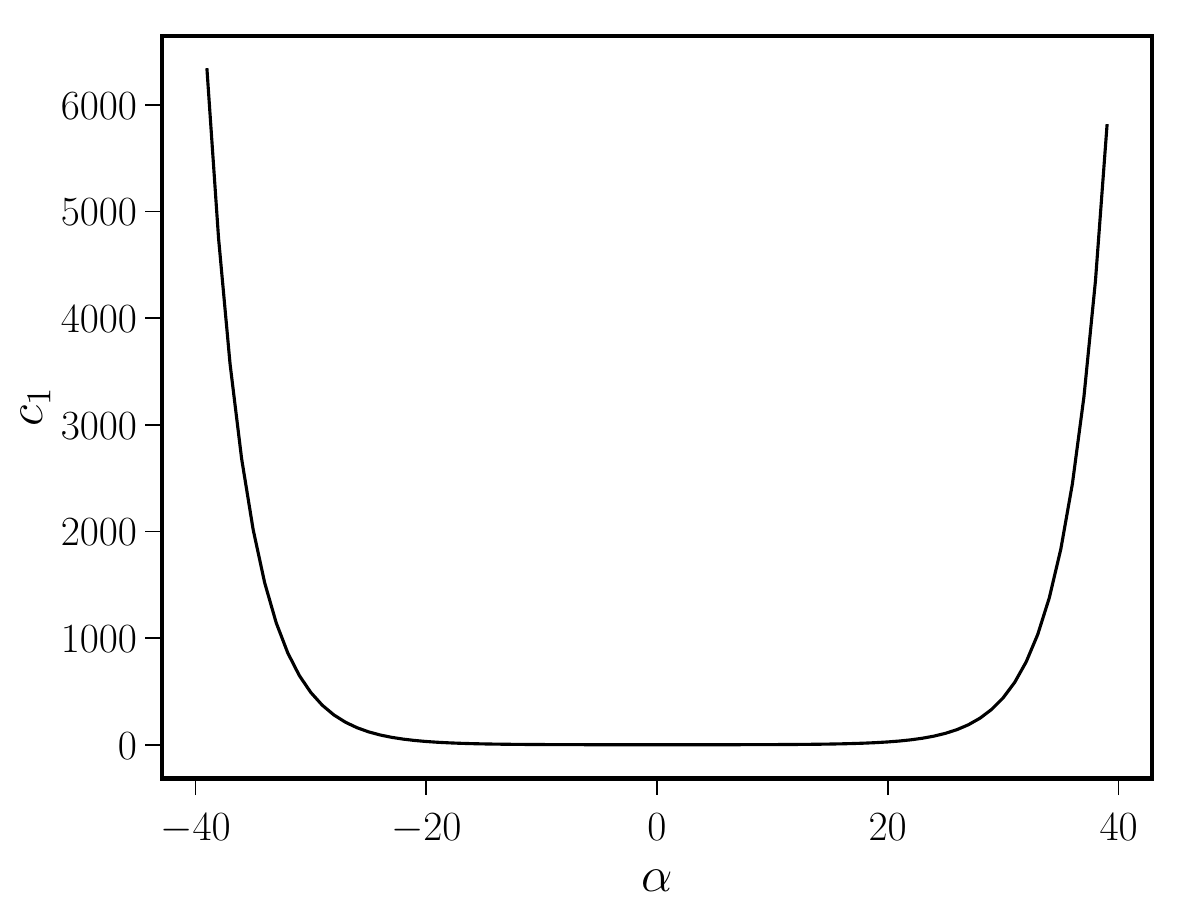}
  \includegraphics[width = 0.48 \textwidth]{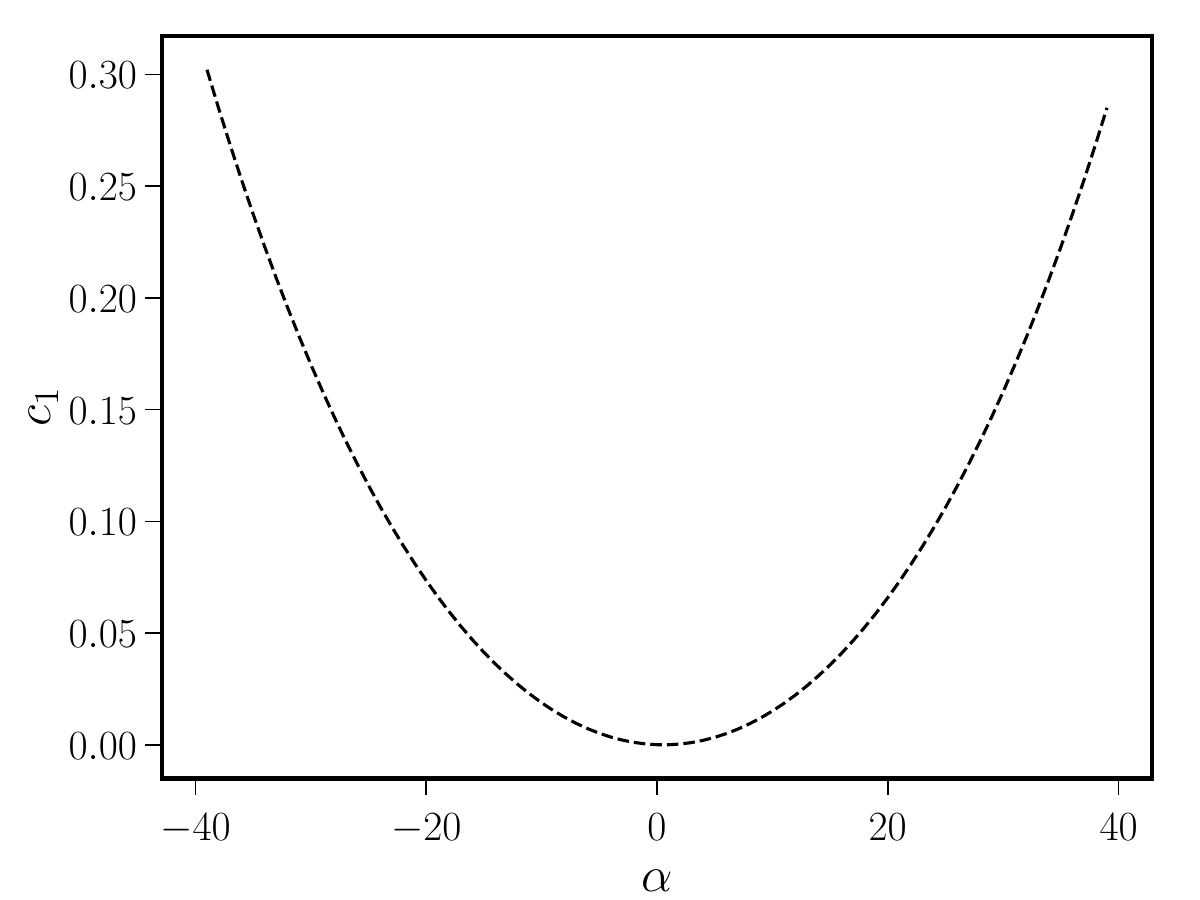}\\
    \includegraphics[width = 0.48 \textwidth]{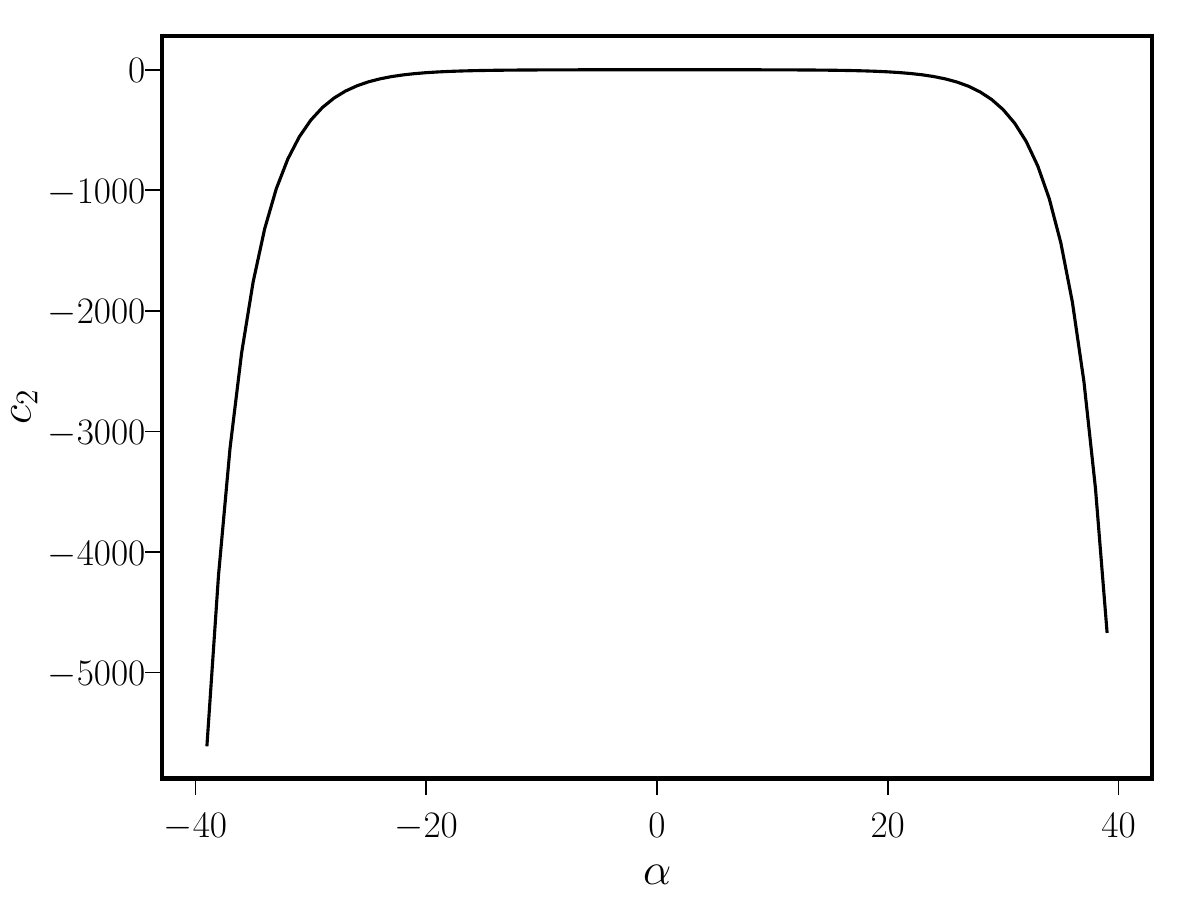}
\includegraphics[width = 0.48 \textwidth]{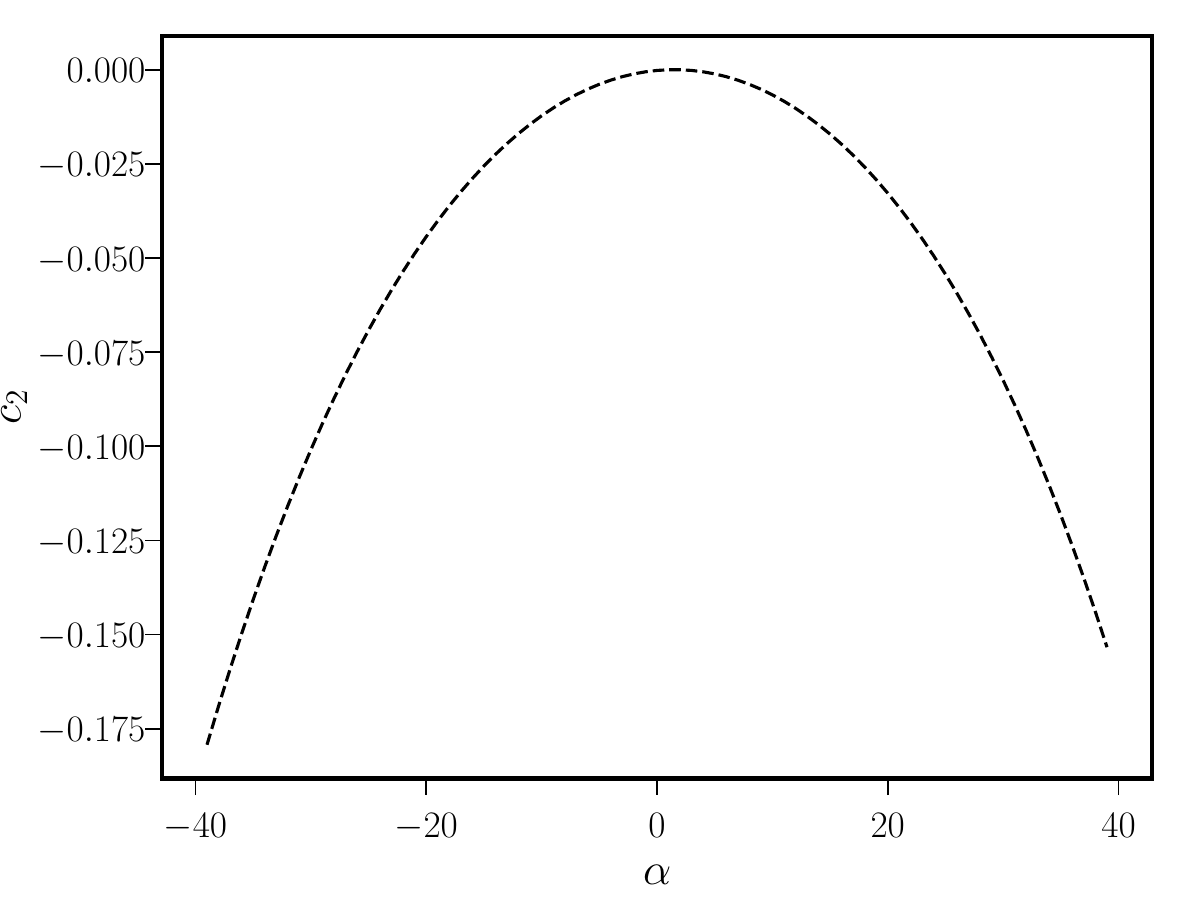}\\
  \includegraphics[width = 0.48 \textwidth]{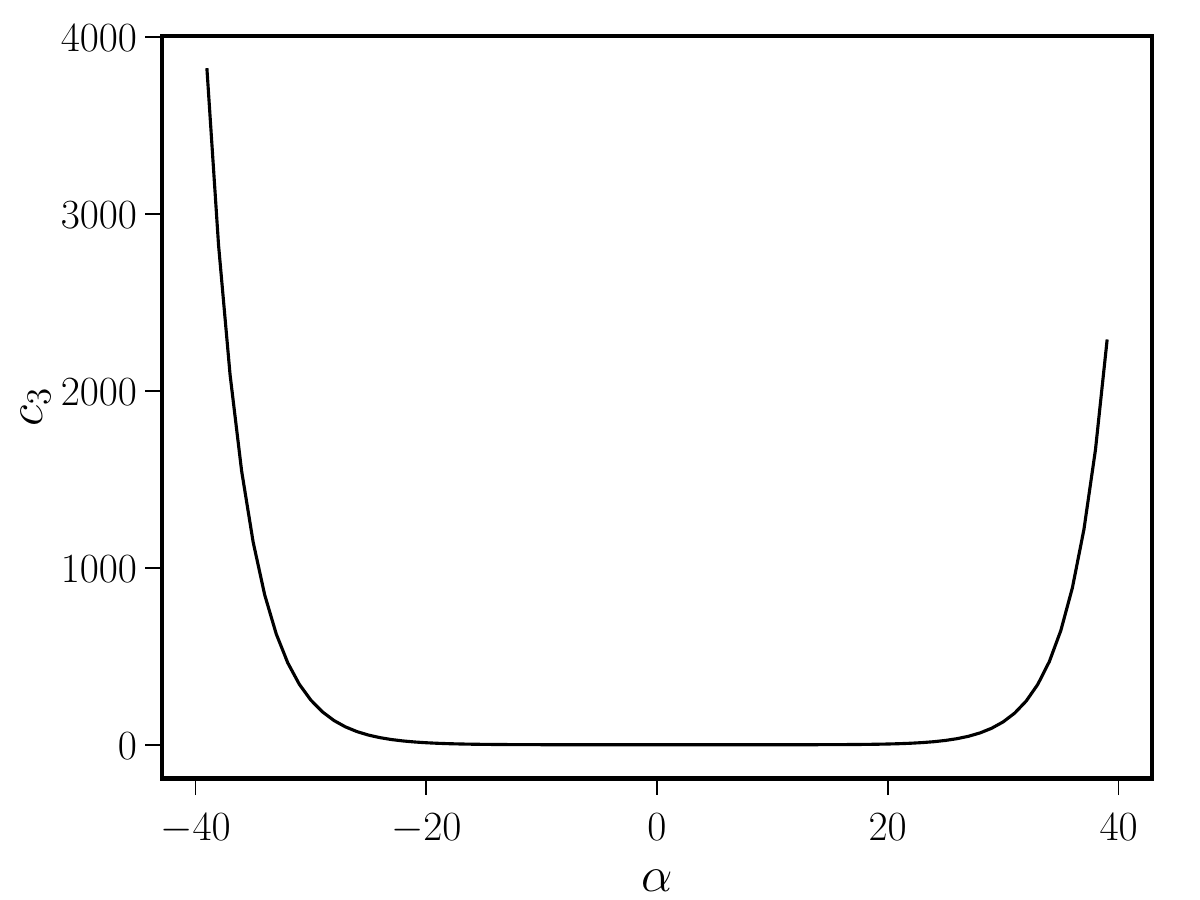}
  \includegraphics[width = 0.48 \textwidth]{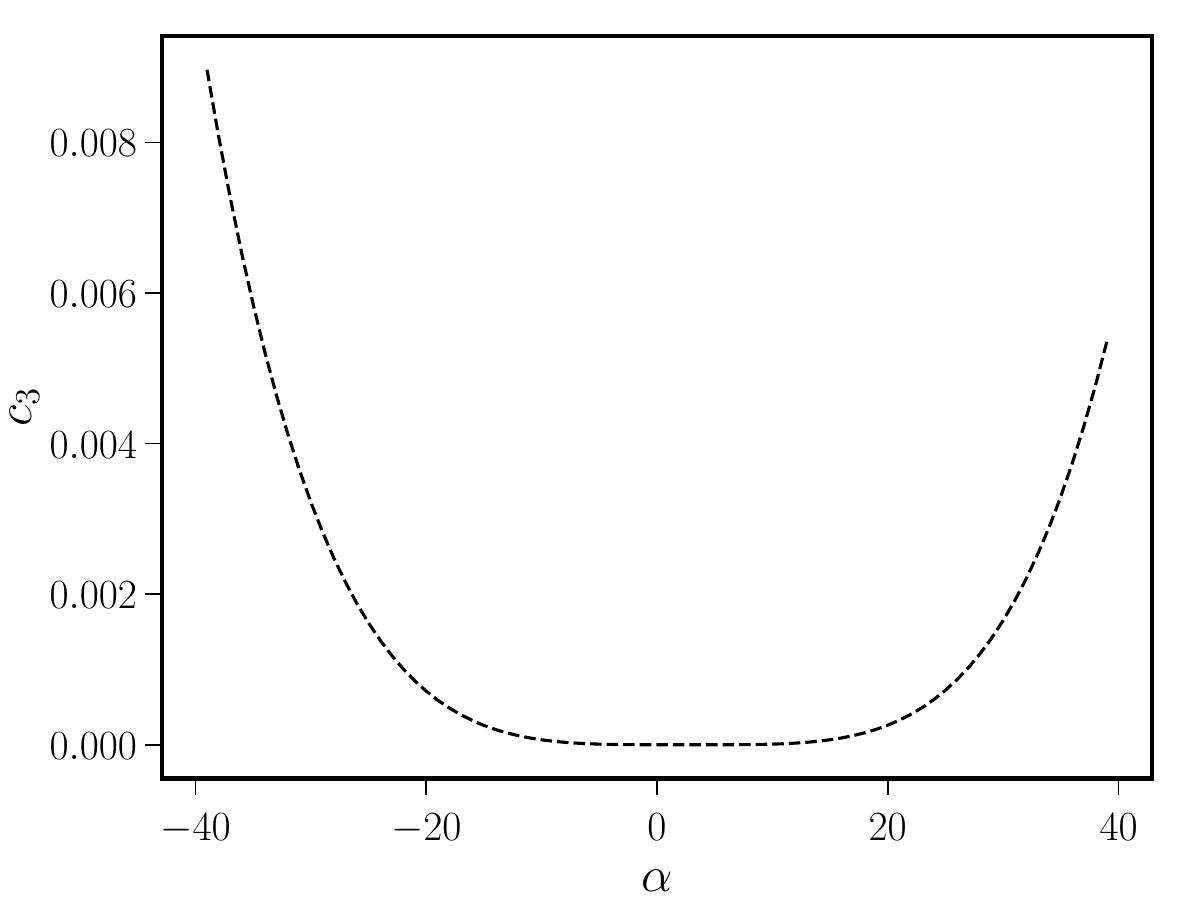}
 \caption{\small The  $c_i$
  for the  power-law Ansatz \eqref{ans_pl1}. {\bf Left plots:} We fix $\beta=0.3$
  and vary the exponent $\alpha$.  {\bf Right plots:}  We fix $\beta=0.04$
  and vary the exponent $\alpha$.
  }
 \label{fig:plot11PL}
\end{figure}

In this power-law scenario, it  is also instructive to investigate how the results
vary with the exponent $\alpha$, while having $\beta$ fixed to a representative value. We plot the 
results in
Fig \ref{fig:plot11PL}, for two choices of the velocity parameter $\beta$, one with $\beta$
large, and one with $\beta$ relatively small. As expected, the amplitude of $c_i$
is quite sensitive to the value of $\beta$. 


The absolute values of the coefficients  $c_i$  of eq \eqref{def_ci}  increase
 as the absolute value of $\alpha$ increases. The larger the exponent, the larger the kinematic effects on the anisotropies. For the coefficient $c_3$, there is a flat plateau 
 around $\alpha=0$, approximately between $-1\le \alpha\le5$.
 Such a flat plateau is present for both large and small values of $\beta$.
   These findings will be useful
 for interpreting the results of  the scenarios we shall discuss next.

 \subsection{\color{black}Second example: a broken power-law SGWB profile} 
\label{sec_bpl}

\begin{figure}[h!]
\centering
  \includegraphics[width = 0.75 \textwidth]{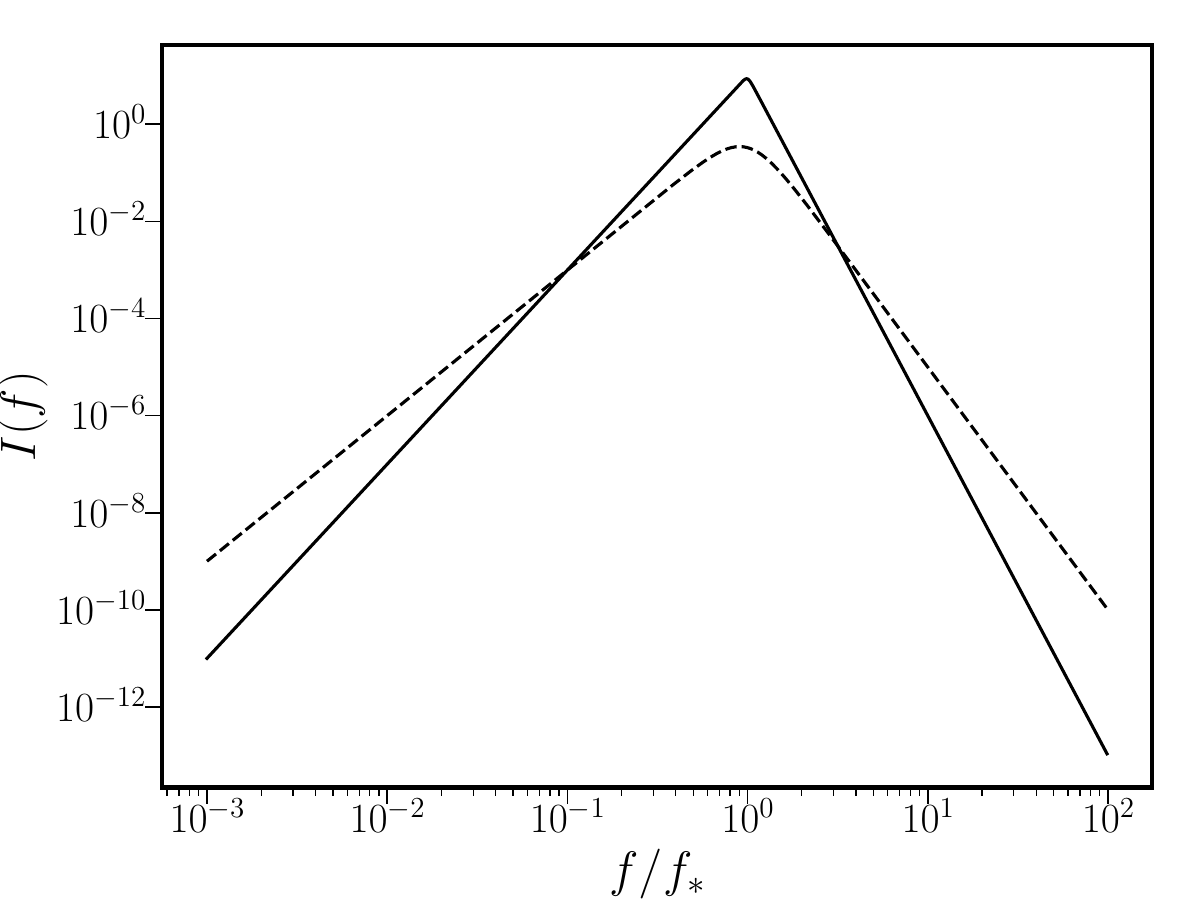}
 \caption{\small Logarithmic plot of  the intensity for  a broken power-law
 profile, eq \eqref{def_inBPL}. {\bf Solid line:} $I_0=10^{-3}$, $\gamma=4$,  $\delta=7$, $\kappa=0.02$,
 $f_{\rm fid}=f_\star/10$. {\bf Dashed line:}  $I_0=10^{-3}$, $\gamma=3$,  $\delta=5$, $\kappa=0.2$,
 $f_{\rm fid}=f_\star/10$.}
 \label{fig:plot1pr}
\end{figure}

We now turn to  a broken power-law profile for the GW intensity  $ {\cal I}(f)$
in the SGWB rest frame. 
 Many examples and realizations of such a profile exist in the literature, see e.g. the survey in \cite{Kuroyanagi:2018csn}. 
In this case, the role of frequency is important, and
kinematic anisotropies can not be factorized
as frequency times direction. We consider the following Ansatz for the SGWB intensity
as function of frequency
  \cite{Sampson:2015ada,Kaiser:2022cma}:
 \be
 \label{def_inBPL}
  {\cal I}^{BPL}(f)\,=\,I_0\,\left(\frac{f}{f_{\rm fid}}\right)^{\gamma} \left[
  1+\left( \frac{f}{f_\star}\right)^{\frac{1}{\kappa}}
  \right]^{-\kappa (\gamma+\delta)}\,.
 \ee
 The exponents $\gamma$ and $\delta$ control the growing and decaying parts of the SGWB frequency
 profiles, respectively. The parameter $\kappa$ controls the smoothness of the transition. The quantity $I_0$ is a normalization factor.  $f_{\rm fid}$
 is a fiducial frequency, and $f_\star$ is a parameter controlling the frequency region where the profile
 changes slope. In Fig \ref{fig:plot1pr} we have plotted the SGWB intensity for concrete examples for
 representative choices of parameters.

\begin{figure}
\centering
  \includegraphics[width = 0.48 \textwidth]{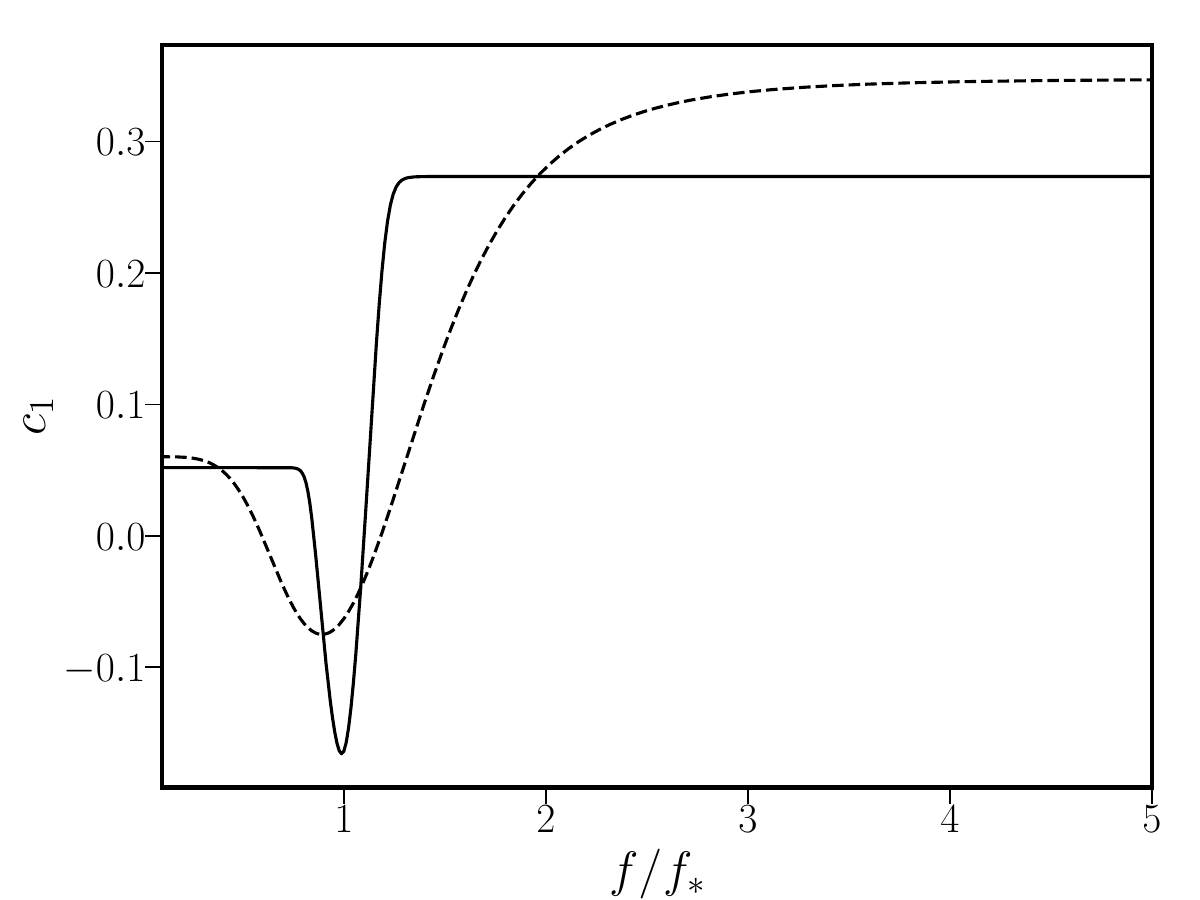}
  \includegraphics[width = 0.48 \textwidth]{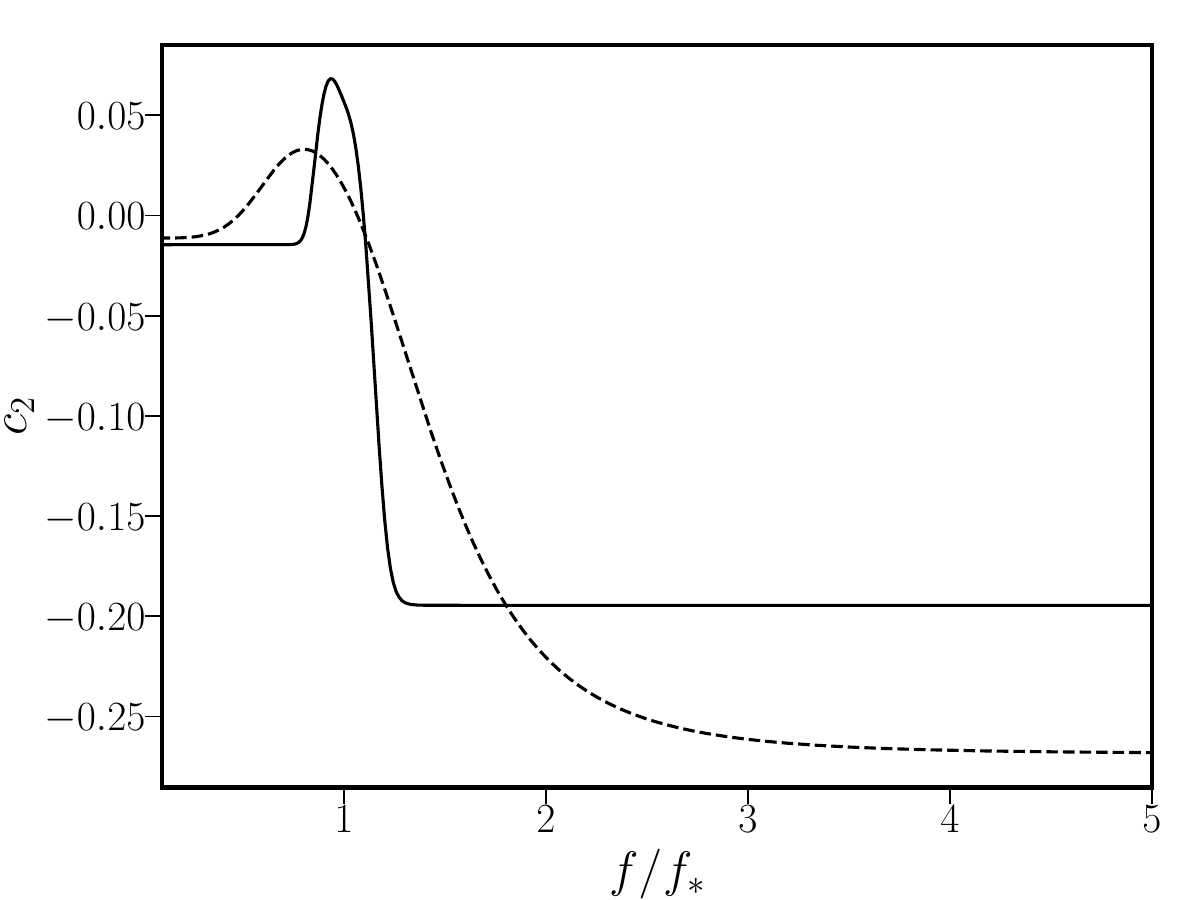}\\
    \includegraphics[width = 0.48 \textwidth]{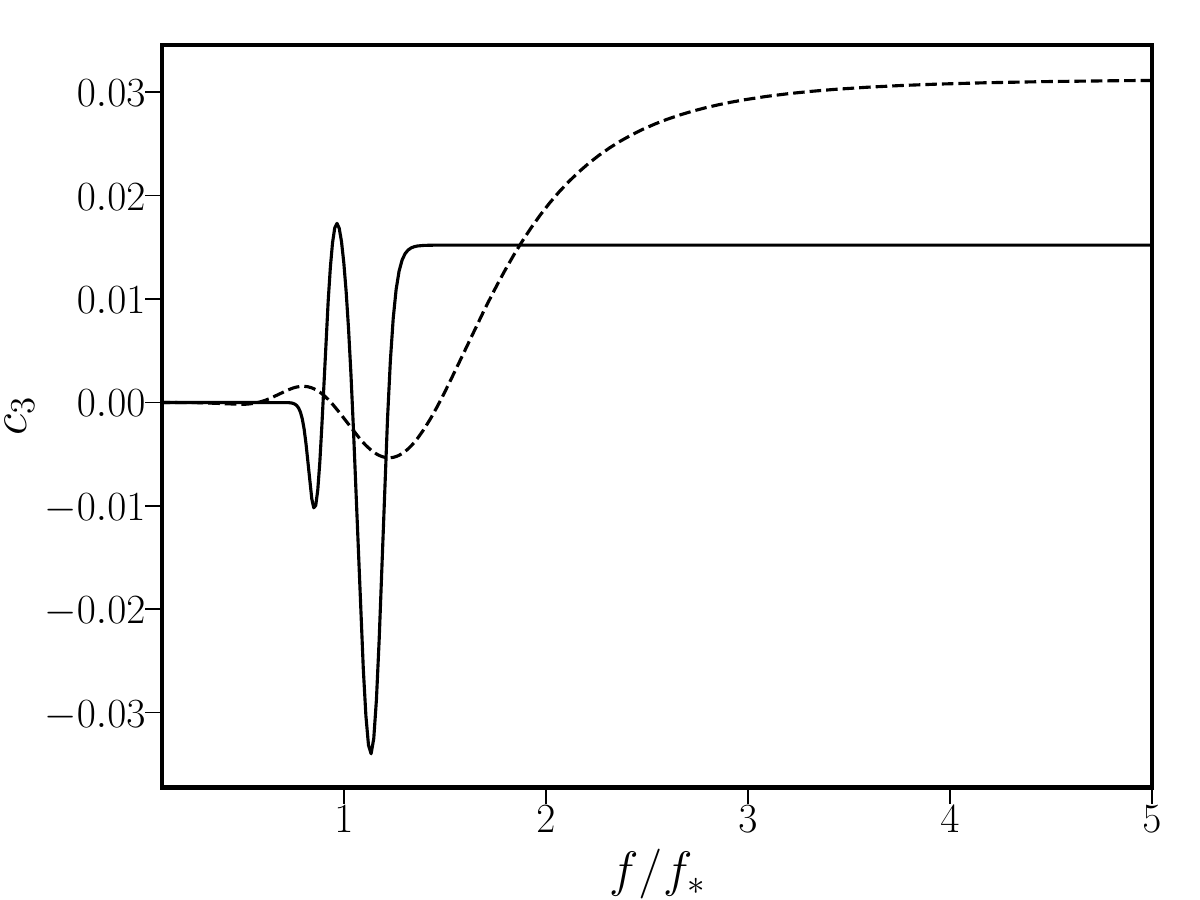}
 \caption{\small The quantities $c_i$
 of eq \eqref{def_ci} 
  for the broken power-law Ansatz \eqref{def_inBPL}. {\bf Solid lines:}    $\beta=0.2$,  and the same parameters as the solid line plot
  of Fig \ref{fig:plot1pr}. {\bf Dashed lines:}    $\beta=0.3$,  and the same parameters as the dashed line plot
  of Fig \ref{fig:plot1pr}.}
 \label{fig:plot2freBPLaa}
\end{figure}

For the case of broken power-law, the quantities   in eq \eqref{def_intK} need to be integrated numerically. The 
coefficients $c_i$ in eq  \eqref{res_OOp} explicitly depend 
on frequency. In fact, we expect them 
to be constant in the frequency ranges corresponding to a constant slope 
 -- in the regions
of growth and decay of the intensity profile -- following the behaviour
described in section \ref{sec_firex}. 
 Their
  non-trivial frequency
dependence is  
  amplified as $\beta$ increases.
 The frequency profiles of the plots in
 Fig \ref{fig:plot2freBPLaa} confirm these expectations.
 Notice that, for the choice of parameters corresponding to the dashed-line
 plot on the right panel of Fig \ref{fig:plot1pr}, the size  of the quantity $c_3$
 in Fig \ref{fig:plot2freBPLaa} is one order of magnitude smaller  than $c_{1,2}$. This 
 is due to the fact that the growing and decaying slopes of the corresponding ${\cal I}(f)$ have
 been chosen to lie in the flat plateau of Fig \ref{fig:plot11PL} (right panel). We will reconsider
 this case in section \ref{sec_snr}.

Furthermore,
we also expect that the smaller the parameter $\kappa$ is in our Ansatz
\eqref{def_inBPL}, the sharper the transition among the growing
and decaying regions of the intensity profile will be. Therefore,
when $\kappa$ is small, the features
in the detector response  as a function of frequency are further
enhanced around $f_\star$. Fig \ref{fig:plot2freBPLaa} confirms these expectations. Notice also that,
for sharp transitions, the absolute value of the amplitude of the $c_i$ around the transition can be larger
than their value in the constant-slope regions (see e.g. fig  \ref{fig:plot2freBPLaa}, upper right panel). This
indicates that the response function can be sensitive to sudden changes in slope, and kinematic anisotropies can be an indicator of such features.

 \subsection{\color{black}Third example: double broken power-law, and resonance}
\label{sec_reso}

As a last example, we consider a double power-law profile
for the SGWB intensity ${\cal I}(f)$. Such a
possibility is physically motivated by early-universe scenarios
in which a SGWB is induced at second 
order in perturbations by a scalar  power spectrum
with a pronounced peak \cite{Ananda:2006af,Baumann:2007zm,Saito:2008jc,Saito:2009jt}. Such models are frequently investigated
in the context of primordial black hole production from inflation -- 
see e.g. \cite{Domenech:2021ztg} for an exhaustive review. Interestingly, if the source
scalar peak is sufficiently narrow, the induced SGWB profile has an 
initial bump, followed by a pronounced, narrow resonance. The details of the bump and of the resonance depend
on properties of the source curvature spectrum, as well as on the
underlying cosmological expansion. Nevertheless, analytical
formulas are available for a number of examples \cite{Espinosa:2018eve,Kohri:2018awv,Pi:2020otn}. Effects
of Doppler anisotropies in these
scenarios have been recently investigated in \cite{Cusin:2022cbb}, in the small $\beta$
limit. 

\begin{figure}[h!]
\centering
  \includegraphics[width = 0.75 \textwidth]{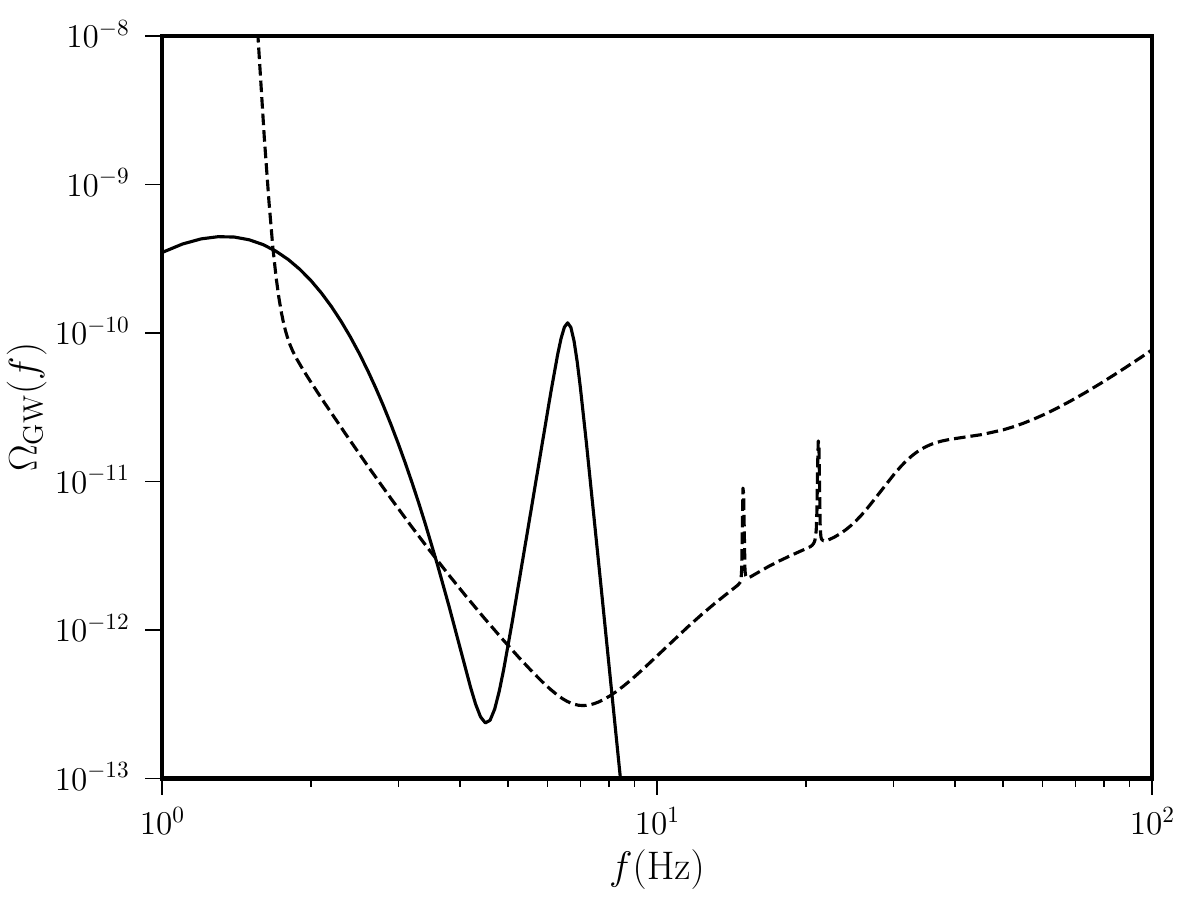}
 \caption{\small Logarithmic plot of  the GW energy density for  a double broken power-law
 profile, eq \eqref{def_inBPL2} (converted to $\Omega_{\rm GW}$ using
 eq \eqref{def_Ois}). We have chosen the following values for the parameters: $I_1\,=\,9 H_0^2 \times10^{-11}/(8 \pi^2)$,
 $I_2 =45 H_0^2 \times10^{-13}/(8 \pi^2)$,
  $\gamma=4$, $\delta=25$
 $\kappa=0.5$, $\gamma_1=20$, $\delta_1=33$,
 $\kappa_1=0.02$, $f_{1}=1/3$, $f_\star=10/3$,
 $f_2=5$, $f_3=20/3$ Hz. A mild bump is followed by a pronounced peak. In the dashed
 line we  plot the Einstein Telescope sensitivity curve $\Sigma_{\rm GW}(f)$ to a stochastic background, 
 with one year of data collection. See the discussion around eq \eqref{def_Som} for explanations.
 }
 \label{fig:plot2DBPL}
\end{figure} 

For
simplicity,  we model such a scenario in terms of double power-law,
essentially duplicating the Ansatz of section \ref{sec_bpl}: 
\be
{\cal I}^{DPL}(f)\,=\,
I_1\,
\left(\frac{f}{f_{1}}\right)^{\gamma-3} \left[
  1+\left( \frac{f}{f_\star}\right)^{\frac{1}{\kappa}}
  \right]^{-\kappa (\gamma+\delta)}+
  I_2
  \,\left(\frac{f}{f_{\rm 2}}\right)^{\gamma_1-3} \left[
  1+\left( \frac{f}{f_3}\right)^{\frac{1}{\kappa_1}}
  \right]^{-\kappa_1 (\gamma_1+\delta_1)}\,,
  \label{def_inBPL2}
\ee
where $I_{1,2}$ are normalization factors,
$f_{1}$, $f_2$, $f_3$ are fiducial frequencies, and $f_\star$
a characteristic frequency around which the first bump occurs. The exponents
have the same roles as in the single broken power-law case (see comments
after eq \eqref{def_inBPL}),
 controlling the slope of the spectrum. See Fig \ref{fig:plot2DBPL}
for a
 phenomenological profile with the desired features. 
 The position of the resonance has been chosen to lie within the best
sensitivity region for the ET-D nominal configuration.

\medskip
\begin{figure}[h!]
\centering
  \includegraphics[width = 0.48 \textwidth]{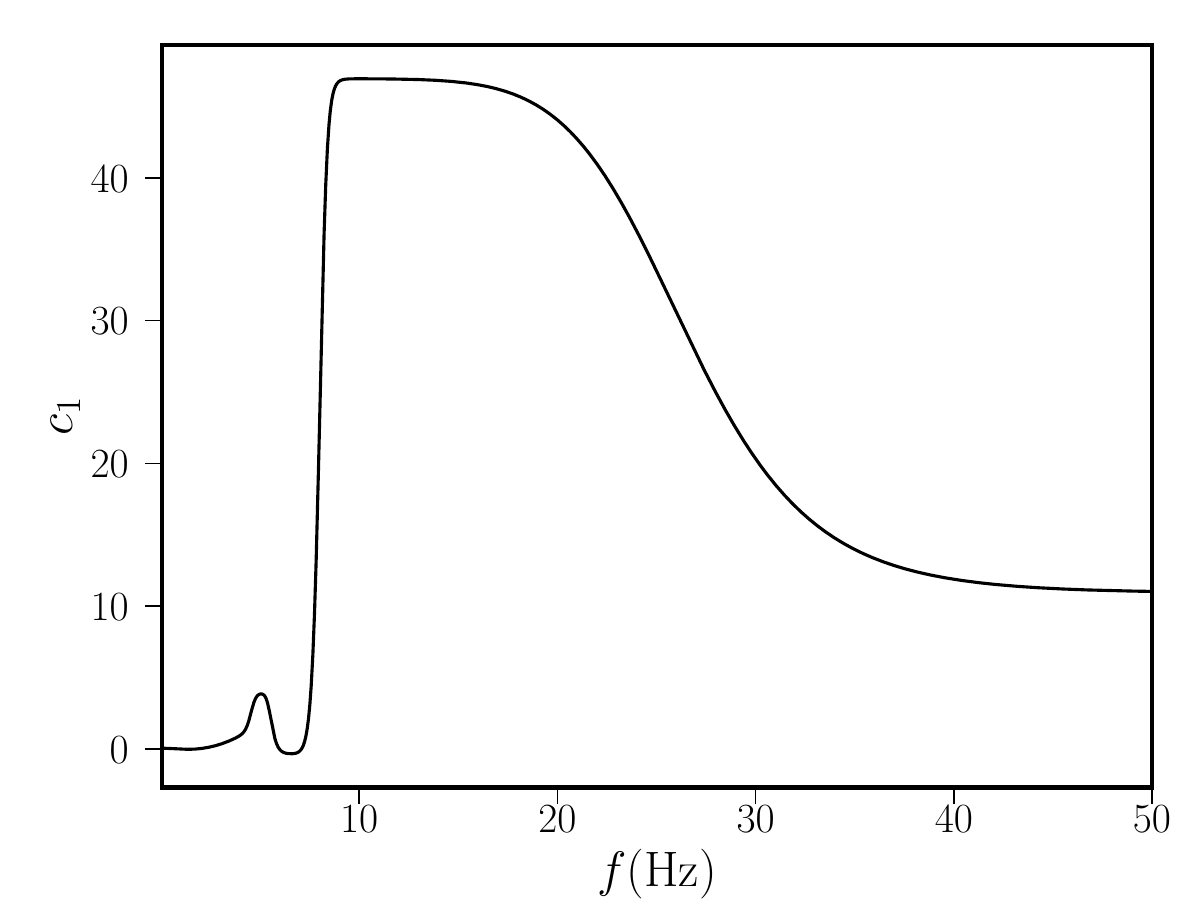}
    \includegraphics[width = 0.48 \textwidth]{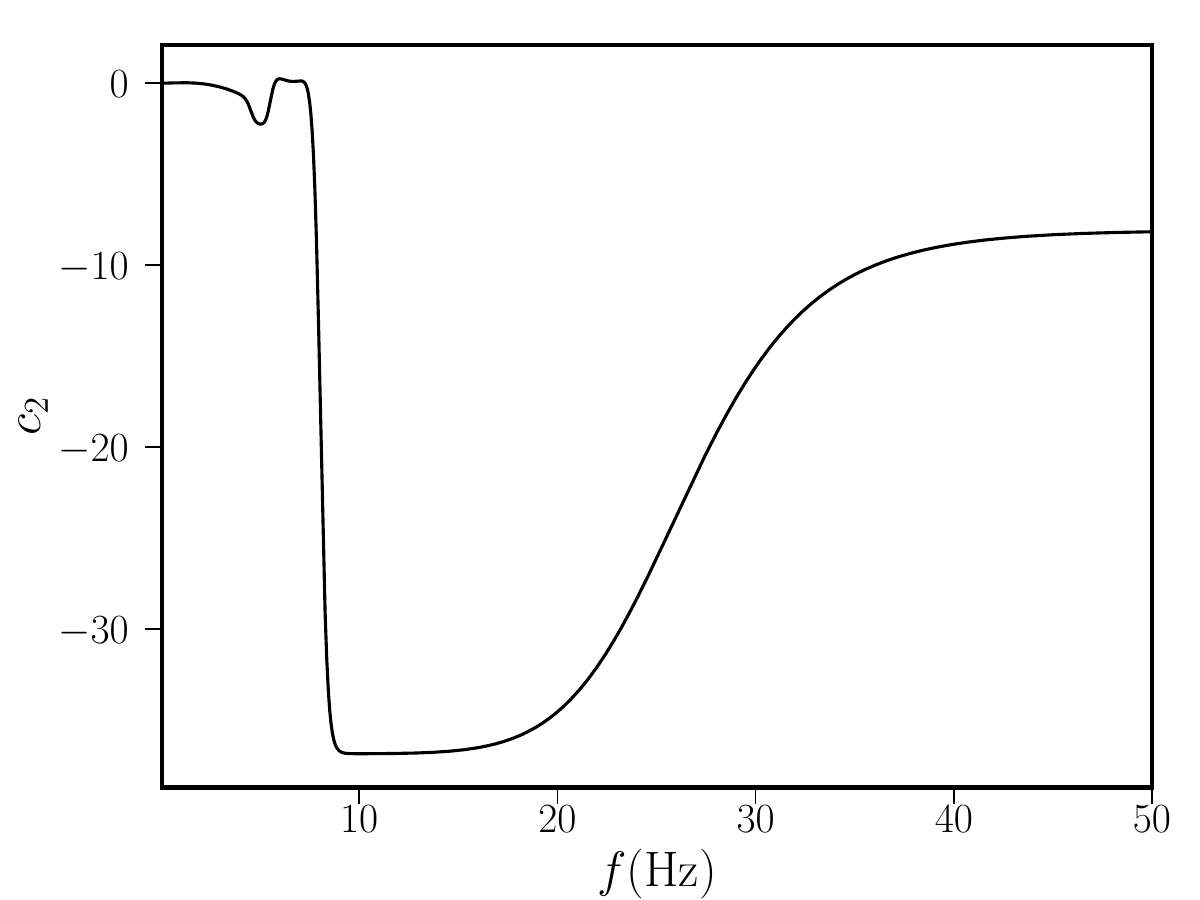}\\
  \includegraphics[width = 0.48 \textwidth]{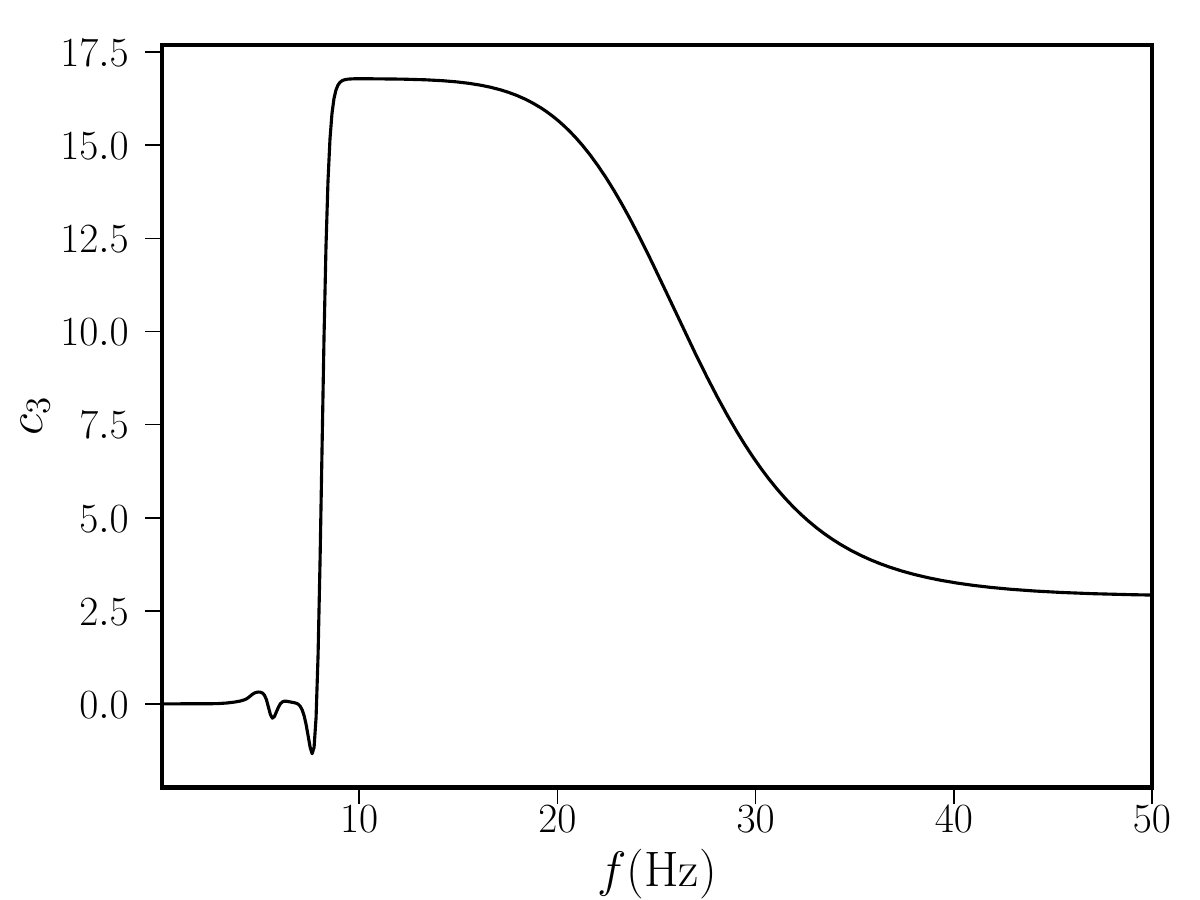}
 \caption{\small The quantities $c_i$
 of eq \eqref{def_ci} 
  for the double broken power-law Ansatz \eqref{def_inBPL2}. 
  We choose $\beta=0.2$, and
  the remaining  parameters are the same as in   Fig \ref{fig:plot2DBPL}.   
  }
 \label{fig:plot2freBPL}
\end{figure}

With such a rich frequency dependence of the initial
spectrum, we can also expect a rich 
frequency profile of the anisotropy parameters
$c_i(f)$ of eq \eqref{res_OOp}.  In proximity of the resonance (the second peak) we expect drastic changes in the amplitude of the $c_i$ as a function of frequency, since the slopes of the intensity ${\cal I}(f)$
(or the GW density) reach large values.  This expectation is confirmed by our results in Fig \ref{fig:plot2freBPL}. 
   In fact, for large slopes, the   absolute value of the $c_i$ become large (see Fig \ref{fig:plot11PL}): this property can enhance the prospects of detectability, as we will learn in section \ref{sec_repex}.  In comparison, the initial mild bump, at frequencies smaller than the resonance
peak, produces  small oscillatory effects. Notice that Fig \ref{fig:plot2freBPL} indicates that  the absolute values of the $c_i$ get larger for an intermediate frequency band, before stabilizing to constant values. The values of the  $c_i$ are following the slopes of the spectrum as it increases, and then decreases, around the resonance region.

The general formulas we developed in section \ref{sec_kin} can also be applied to any further physically
motivated Ans\"atze for ${\cal I}(f)$. In fact, it  would be interesting to carry out a more systematic  investigation of the
 ET response function to kinematic anisotropies for a  greater variety of frequency
 profiles. 
 If any of the features associated with  the kinematic anisotropy
parameters can be detected,
they might represent a further indirect probe of the frequency profile of ${\cal I}(f)$, besides  direct methods
 \cite{Caprini:2019pxz}. 
   We plan to investigate these subjects  in future works.

 \section{\color{black}Detectability  and signal-to-noise ratio}
 \label{sec_snr}
 
 In this section we investigate the prospects of detectability
 of kinematic anisotropies by means of ground-based
 interferometers, focussing on the Einstein Telescope. Our aim is to determine an optimal
 estimator for a quantity sensitive to kinematic anisotropies, 
 study the corresponding signal-to-noise ratio, and consider
 some representative examples of non-monotonic frequency profiles. 
 
 We make the hypothesis that
 noise and 
  GW signal  (in the SGWB rest frame)  are
 stationary. But recall that, as we learned in section \ref{sec_kin},
 a feature of the ET response function
 to kinematic anisotropies is its {\it time-dependence}. The
 orientation
 of the detector with respect to the velocity vector between the ET and SGWB
 frames changes with time, following the daily and annual motions
 of the Earth. Such time-dependence of the signal is precisely the key
 for determining an optimal estimator sensitive to kinematic anisotropies. 
 
 \subsection{\color{black}Disentangling the signal time-dependence}
 \label{sec_disen}
 
  Inspired by the works \cite{Allen:1996gp,Mentasti:2020yyd},  we start discussing a method to disentangle
 the daily time-dependence of the signal, and formulate time-independent
 quantities which are easier to deal with. So far,
  our results have been presented in a  covariant form: see e.g. eq \eqref{res_OOp}.
  To proceed, we  choose a  convenient reference frame. Let our reference
 system be anchored to the Earth, with $\hat z$ axis along the earth rotation axis. 
The detector tensors $d^{ab}$ in  eq \eqref{res_OOp} are constants, and what varies
with time is the velocity vector $\hat v$. For any given time $t$,  we can split the velocity vector into two parts - one
along the $\hat z$ axis, and the other perpendicular to it:
\bea
\hat v &=&\left( \hat v \cdot \hat z \right)\,\hat z
+
\left[
 \hat v-  \left( \hat v \cdot \hat z \right)\,\hat z
 \right]\,.
\eea

The vector component parallel to $\hat z$, which we dub $\vec v_\parallel$, does not change
with time, being along the Earth rotation axis. The vector component orthogonal to $\hat z$,
which lies on the plane $(\hat x, \,\hat y)$, undergoes  a sinusoidal
daily modulation with period $T_e\,=\,24$ hours. Dubbing  $\bar f_e\,=\,1/T_e$ the frequency of the Earth rotation, 
and indicating with ${\vec v}_{\perp}\,\cos{\left(2 \pi\,\bar f_e\,t\right)}$ the time-dependent  component
of the velocity vector in the plane $(\hat x, \,\hat y)$, we can write
\bea
\label{def_spv}
\hat v (t)&=&{\vec v}_\parallel+{\vec v}_{\perp}\,\cos{\left(2 \pi\,\bar f_e\,t\right)}\,,
\eea
with ${\vec v}_\parallel$, ${\vec v}_{\perp}$ constant vectors.

Using the split of eq \eqref{def_spv},  we can decompose the response
function  $ {\cal R}_ {{\cal O},\,{\cal O}'}$ of eq \eqref{res_OOp}
into a finite set of  terms, each one with its own dependence on time:
\bea
\label{def_Rmmp}
  {\cal R}_ {{\cal O}\,{\cal O}'}( f, t ,t')&=& \sum_{m, m'=-2}^2 {\cal R}_ {{\cal O}\,{\cal O}'}^{(m,m')}(f)\,
e^{2 \pi i\, \bar f_e (m t+m' t')}\,.
\eea
 The time-independent (but frequency-dependent)  ${\cal R}_ {{\cal O}\,{\cal O}'}^{(m,m')}(f)$  coefficients are even under interchanges  of $m\to-m$, $m'\to-m'$. Moreover, they have the property ${\cal R}_ {{\cal O}\,{\cal O}'}^{(m,m')}\,=\,{\cal R}_ {{\cal O}'\,{\cal O}}^{(m',m)}$. 
  They can be found   in appendix \ref{app_Rmmp}, expressed in terms of
contractions of detector tensors with the vectors $\vec v_{\parallel,\,\perp}$.  Since
eq \eqref{res_OOp}  contains three contributions only, the sum in eq \eqref{def_Rmmp} spans
only a finite number of terms. 

  \subsection{\color{black}Defining an optimal estimator}

  To continue, we follow \cite{Allen:1996gp,Mentasti:2020yyd},  introducing the notion of time-dependent
  Fourier transform as  
\be
\tilde \Phi_{\cal O}(t,f )\,\equiv\,\int_{t-\tau/2}^{t+\tau/2}\,d t'\,e^{-2\pi i\,f t'}\,\, \Phi_{\cal O}(t')
\,,
\ee
with $\tau$ being a convenient chopping  time much longer than the time spent by light in
travelling among different parts of the interferometer system (so as to ensure the
signal develops correlations), but much shorter than the daily period of the Earth (so that
the signal can be taken as  constant during the interval $\tau$). 

We define the quantity ${\cal C}(t)$ as

\bea
{\cal C}(t)\,\equiv\,\sum_{{\cal O} {\cal O}'}\,\int_{-\infty}^{\infty} d f\,\,\tilde Q_{{\cal O} {\cal O}'}(f)\,\tilde \Phi_{\cal O}(t,f )\,\tilde \Phi^*_{\cal O'}(t,f )
\,,
\eea
and use it as the estimator of kinematic anisotropies. 
The function  $\tilde Q_{{\cal O} {\cal O}'}(f)$ is  the optimal filter to be determined. 
We
disentangle the Earth rotation effects  in the estimator by Fourier expanding:
\be
{\cal C}(t)\,=\,\sum_{m}\,{\cal C}_m e^{2 \pi i\,  m\,\bar f_e\,t}\,.
\ee 
Hence, by inversion, the time-independent coefficients ${\cal C}_m$ are given by
\be
{\cal C}_m \,=\,\frac{1}{T} \int_0^Td t\,{\cal C}(t)\,e^{-2 \pi i  m\,\bar f_e\,t}\,,
\ee 
with $T$ being the total time of data collection, which we assume to be 
a multiple of $T_e$. 
The ${\cal C}_m $ are the constant quantities we are interested
in for estimating the detectability of the signal. The corresponding SNR for each index $m$ is
defined as 
\be
\label{def_snrDEM1}
{\text{SNR}}_m\,=\,\frac{\langle {\cal C}_m \rangle}{\langle {\cal C}^2_m \rangle^{1/2}}\,.
\ee
We derive the expression for the optimal ${\text{SNR}}_m$ in the technical appendix
\ref{app_snr}.
 The crucial property  we will use  is that, for non-vanishing $m$, only the signal
 contributes to the numerator of eq \eqref{def_snrDEM1}, since the noise is
 stationary, and its contribution cancels when it appears within oscillatory integrals. This is why
 we can exploit the daily 
  time modulation of the signal for extracting information on kinematic anisotropies.
  
  The result is
 \be
 \label{def_snrDEM}
{\text{SNR}}_m\,=\,\sqrt{2 T}\,
\left( 
\int_{0}^\infty d f\,
{\cal S}_m^2(f)\,\frac{{\cal I}^2 (f)}{N^2(f)}
 \right)^{1/2}\,,
 \ee
for $m\neq0$, under the hypothesis of common noise $N$ for any non-null channel. We introduced the combination of response functions
\be
\label{def_Sm}
{\cal S}_m(f)\,=\,
\left|
\sum_{{\cal O} {\cal O}'} 
\sum_{m', m''=-2}^2\,\delta_{K}(m-m'-m'')
 {\cal R}_{{\cal O} {\cal O}'}^{(m',\,m'')}(f)
\right|\,,
\ee
with $\delta_K$ being the Kronecker delta. We learn that
the optimal SNR depends on the frequency-dependent quantities $ {\cal R}_{{\cal O} {\cal O}'}^{(m-m',\,m')}(f)$,
associated with the Doppler anisotropies of the SGWB signal.

  \subsection{\color{black} Representative examples }
\label{sec_repex}

\begin{figure}[h!]
\centering
  \includegraphics[width = 0.75 \textwidth]{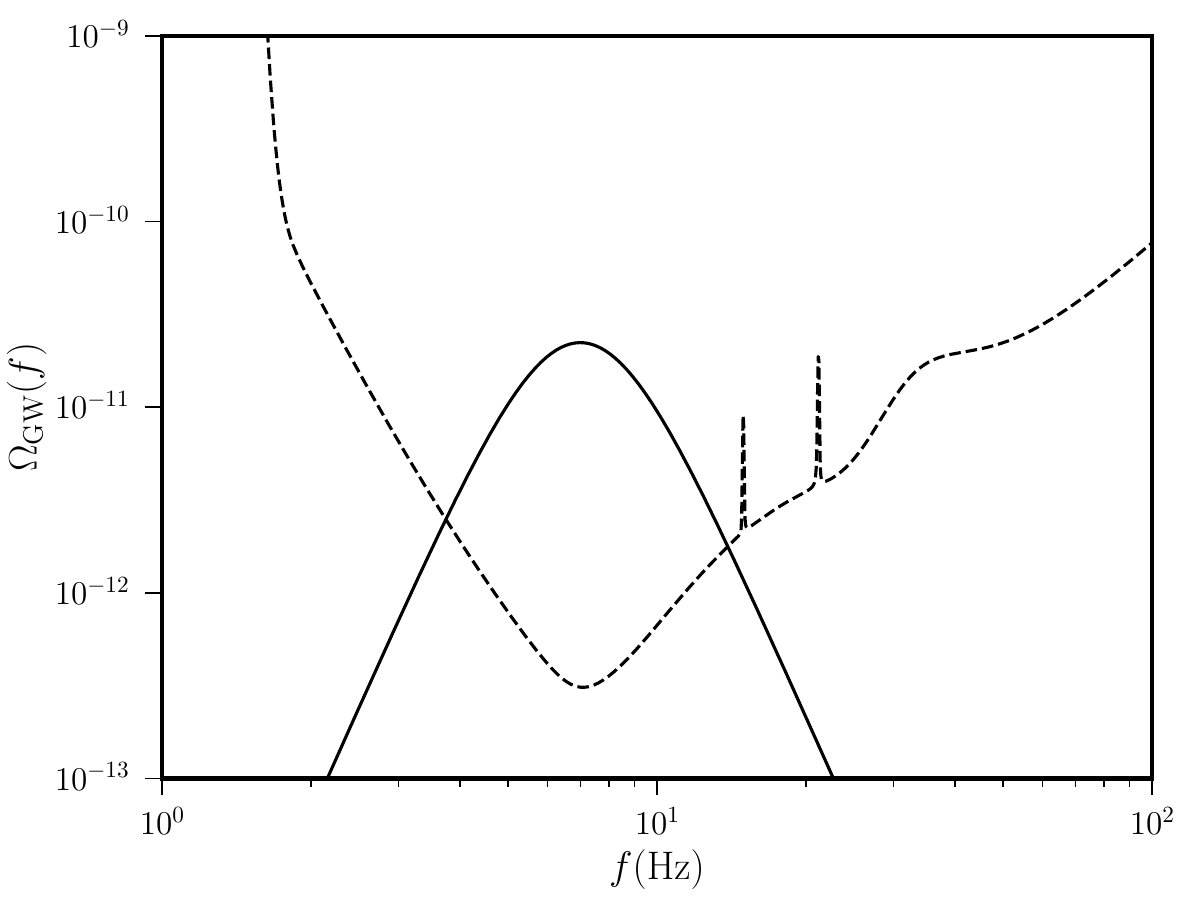}
 \caption{\small 
 Logarithmic plot of  of  the GW energy density  for  a broken power-law
 profile, eq \eqref{def_ex1a}.
 $\Omega_{\rm GW}^{(0)}= 10^{-15}$, $\gamma=3$,  $\delta=9$, $\kappa=0.2$,
 $f_{\rm fid}=1$ Hz, $f_\star=7$ Hz.  In the dashed
 line we  plot the  sensitivity curve $\Sigma_{\rm GW}(f)$ of eq \eqref{def_Som}. 
 }
 \label{fig_plot_exam1}
\end{figure}

We now apply  the previous findings to the representative example of  SGWB with a
 broken power-law frequency profile. 
     First, we select the same Ansatz as eq \eqref{def_inBPL} for the GW
 intensity, convert it  to energy density by writing
 \be\label{def_ex1a}
 \Omega_{\rm GW}(f)\,=\, \Omega_{\rm GW}^{(0)}\,f^3\,\left(\frac{f}{f_{\rm fid}}\right)^{\gamma} \left[
  1+\left( \frac{f}{f_\star}\right)^{\frac{1}{\kappa}}
  \right]^{-\kappa (\gamma+\delta)}\,,
 \ee
 with $\Omega_{\rm GW}^{(0)}\,=\,{4\pi^2\,I_0}/{( 3 H_0^2)}$, 
 and represent in Fig \ref{fig_plot_exam1} the profile of interest. We choose a convenient set of parameters
 in which the SGWB profile changes slope at the frequency $f = 7$ Hz corresponding to the maximal sensitivity for ET-D.  
 
\begin{figure}[h!]
\centering
  \includegraphics[width = 0.49 \textwidth]{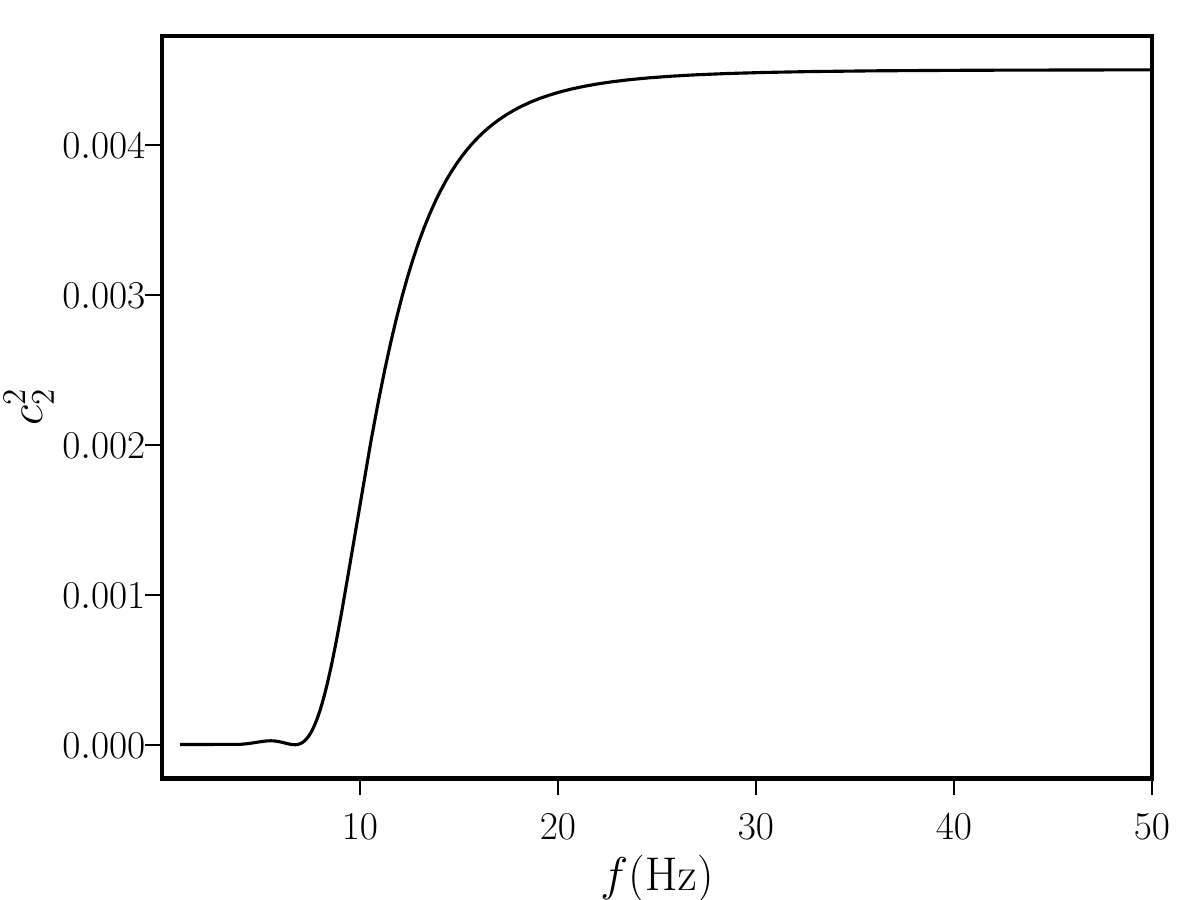}
    \includegraphics[width = 0.49 \textwidth]{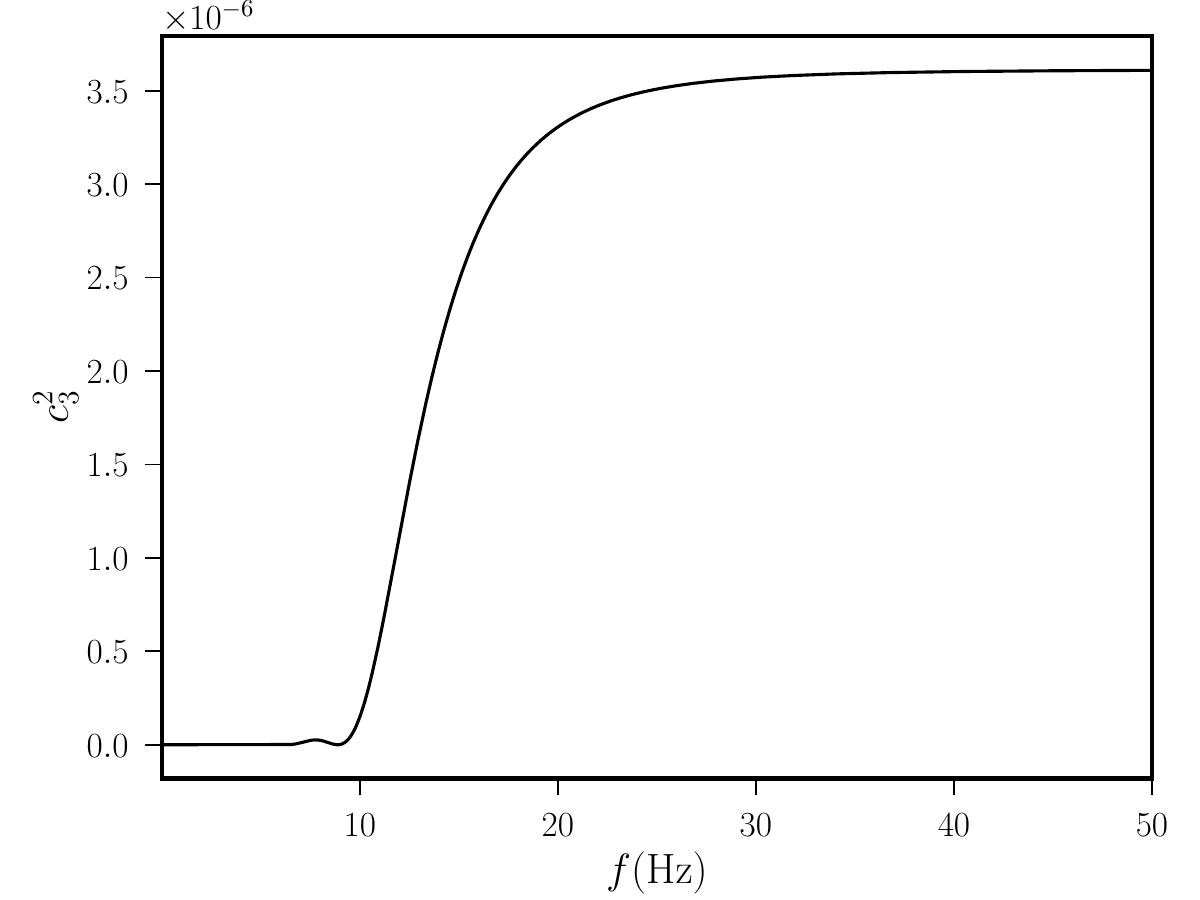}
 \caption{\small The  profiles of $c_2^2$ and $c_3^2$ for the profile  of Figure \ref{fig_plot_exam1}. 
 We fix $\beta=0.1$.
  }
 \label{fig_plot_examc23}
\end{figure}

In this example, the slopes in the growing and decaying phases of the spectrum are small, and consequently 
we expect that the $c_3(f)$ function is much smaller than $c_2(f)$ (see the discussion in sections \ref{sec_firex}
and \ref{sec_bpl}, in
particular figures \ref{fig:plot11PL} and \ref{fig:plot2freBPLaa}).  We confirm this fact by computing  the squares of $c_{2,3}(f)$ for the example at hand. (These are 
quantities we will need in a moment.) 
We plot the result  in Fig \ref{fig_plot_examc23}: manifestly,  the value of $c_3$ is orders of magnitude smaller than $c_2$ over the entire interesting range of frequencies.  Moreover, both $c_2^2$ and $c_3^2$ act as high-pass filters in frequency, being vanishingly small for frequencies smaller than around
$13$ Hz, and almost constant  for frequencies above this value. Since 
both quantities
 enter in the response function and in the expression
for ${\rm SNR}_{m}$, 
such a behaviour makes manifest the importance of frequency
dependence of the  signal ${\cal I}(f)$ for  forecasting the
detectability of kinematic anisotropies.

\bigskip

Depending on the index $m$, by evaluating the quantity ${\cal S}_m$,
we can probe both the profiles for $c_2$ and $c_3$.

Let us start focussing on the
  case $m=2$. The function ${\cal S}_m$  
reads  (we make use of the formulas in Appendix \ref{app_Rmmp})
\be
{\cal S}_2\,=\,\sum_{{\cal O}{\cal O}'}
{\cal R}_ {{\cal O}{\cal O}'}^{(1,1)}\,=\, c_2(f)\,\left( \sum_{\cal O \cal O'} v_{\perp}^a  \,d_{{\cal O}\,ab}\,d^{\,\,\,b}_{{\cal O}'\,d}\,
 v_{\perp}^d
 \right)\,.
\ee
Here we have included only the contribution due to $c_2$, since
the additional contribution of $c_3$, being subdominant, can be neglected (see Fig \ref{fig_plot_examc23}).
The corresponding SNR$_2$ reads
\be
\label{res_snr2}
{\rm SNR}_2 = \left(\sqrt{\frac{T}{1 {\rm year}}}\times \frac{\Omega_{\rm GW}^{(0)}}{{10^{-12}}}\right)\left(\frac{\Sigma_{\mathcal{O}\mathcal{O'}}\,v^a_\perp\,d_{\mathcal{O}\,ab}\,d_{\mathcal{O'}\,d}^b\,v_\perp^d}{\frac{4}{5}\,\Sigma_{\mathcal{O}\mathcal{O'}}\,d_{\mathcal{O}\,ab}\,d_{\mathcal{O'}}^{ab}}\right)\left(\int_{f_{\rm min}}^{f_{\rm max}}\,{\rm d}f\,\mathcal{A}(f)\right)^{1/2},
\ee

with 
\be
\label{def_AF}
\mathcal{A}(f) = c_2^2(f)\,\left(\frac{\Omega_{\rm GW}(f)}{\Omega_{\rm GW}^{(0)}}\right)^2\left(\frac{10^{-12}}{\Sigma_{\rm GW}(f)}\right)^2,
\ee
where we use $1$ year$= 365.25$ days $= 31.56 \times 10^6$  ${\rm Hz}^{-1}$, and we introduce  the
quantity 
 \be
  \Sigma_{\rm GW}(f) = \frac{4\,\pi^2\,f^3}{3\,H_0^2}\left(\frac{N(f)}{\frac{4}{5}\,\Sigma_{\mathcal{O}\mathcal{O'}}\,d_{\mathcal{O}\,ab}\,d_{\mathcal{O'}}^{ab}}\right)\left(\frac{1}{\sqrt{31.56\times 10^6}}\right),
 \label{def_Som}
 \ee 
which corresponds to the sensitivity curve for the detection of a stochastic background with one year of data collection -- see Section 
IVA of   \cite{Smith:2019wny}, and Figure 14 of \cite{Maggiore:2019uih}. 
  Besides the overall $f^3$
 and constant
 factors, this function    is defined in terms of the noise correlation $N$ of eq \eqref{def_noise} (common to all non-null channels,
 and built in terms   of publicly available ET-D specifications~\footnote{\url{http://www.et-gw.eu/index.php/etsensitivities}} \cite{Hild:2010id}), over the response function for an isotropic background (defined as in
 eq \eqref{res_OOp} with all $c_i$ set to zero), and over the square root of one year expressed in Hz$^{-1}$.

\bigskip

The SNR$_2$  of eq \eqref{res_snr2} is then made of three coefficients, with transparent physical interpretations:
\begin{enumerate}
\item The overall square root of the observation time, as expected. \textcolor{black}{We accompany it with the coefficient ${\Omega_{\rm GW}^{(0)}}/{10^{-12}}$, setting the fiducial
overall normalization for the SGWB energy density as in Fig \ref{fig_plot_exam1}}.
\item A purely geometrical contribution, depending on the orientation of the detector with respect to the projection
of the velocity vector normal to the Earth rotation axis.
\item An integral in frequency, which depends on the profile of the SGWB, as well as on the detector specification. The integral
is computed over the detectable frequency band of ET.  It contains the square of the GW density (see Fig \ref{fig_plot_exam1}) and the square of the Doppler
 coefficient $c_2$ (see Fig \ref{fig_plot_examc23}, left panel), that makes it sensitive to the degree of kinematic anisotropy of the scenario. Finally, it also depends on the 
 square of noise energy density (see eq \eqref{def_Som}).
\end{enumerate}

\begin{figure}
\centering
  \includegraphics[width = 0.75 \textwidth]{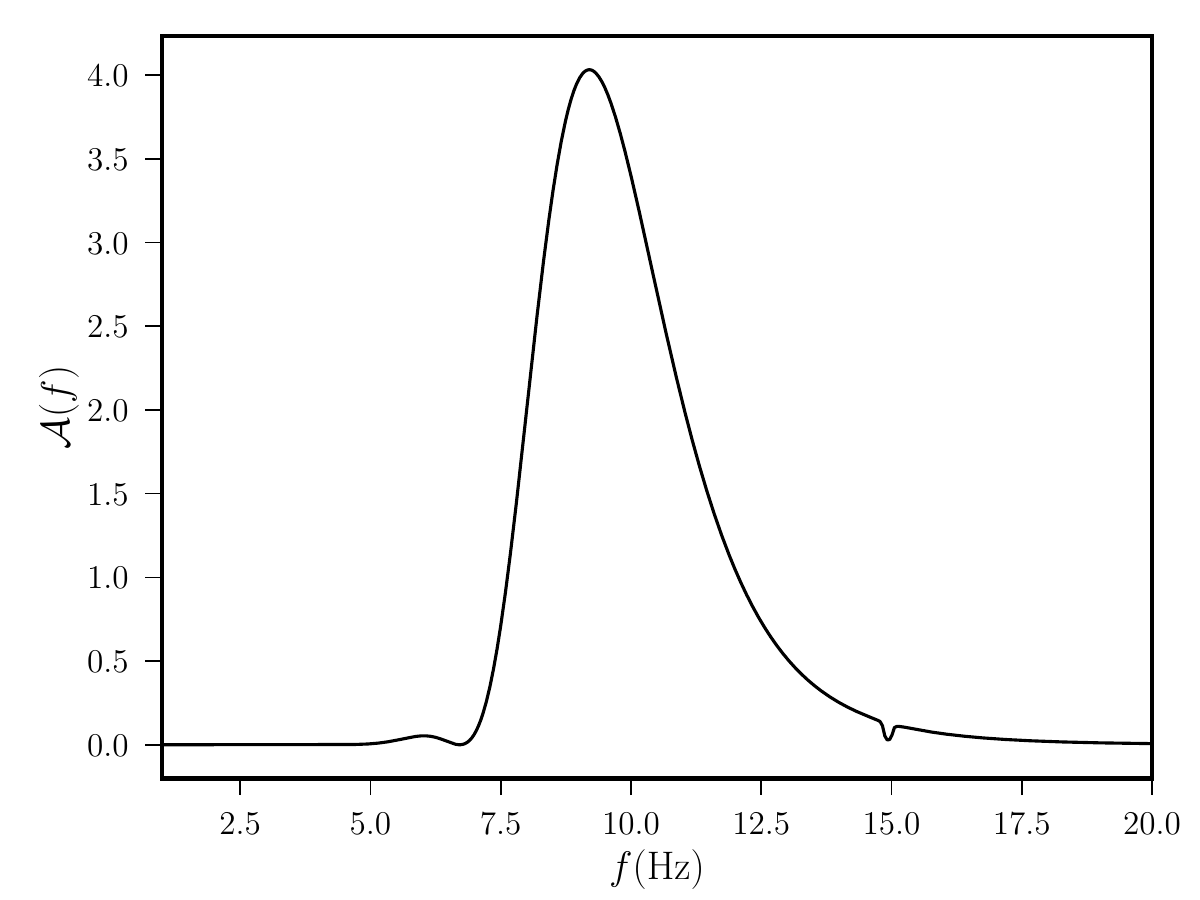}
 \caption{\small{  Plot of the integrand function ${\cal A}(f)$ of eq \eqref{def_AF}, computed for the example 
 of Figs. \ref{fig_plot_exam1},  \ref{fig_plot_examc23}. This function has a peak at frequencies   around $10$ Hz, in proximity of the region of 
 maximal sensitivity of ET.
   }}
 \label{fig:plotSNRf}
\end{figure}

For definiteness, in Fig \ref{fig:plotSNRf} we represent the  function  ${\cal A}(f)$ of eq \eqref{def_AF}, which appears
in the frequency integral in eq \eqref{res_snr2}. We notice a very pronounced peak at frequencies around $10$ Hz, nearby the maximal sensitivity of the instrument. It also  corresponds to the region where
the SGWB changes slope. Notice that the peak is slightly slanted towards high frequencies: we interpret this behaviour to be due 
to the `high-pass' filter function $c_2^2$, as commented at the beginning of this subsection.

The integral over frequencies corresponding to the last factor of eq \eqref{res_snr2} can be numerically computed. We find
\be
\label{res_int1}
\left( 
\int_{1\,{\rm Hz}}^{10^4\,{\rm Hz}} \frac{d f}{\rm 1\,Hz}\,
 {\cal A}(f)
 \right)^{1/2}\,=\,3.66\,.
\ee 
\textcolor{black}{Hence we learn that, if the geometrical second  factor of eq \eqref{res_snr2}  is of order one,  we need an ${\Omega_{\rm GW}^{(0)}}\simeq{10^{-12}} $ for ensuring ${\rm SNR}_2>1$}. The integral over frequencies  in eq \eqref{res_int1} helps in increasing the prospects
of detectability - a feature that has been exploited in the past in the context of power-law sensitivity curves \cite{Thrane:2013oya} (see also
\cite{Chowdhury:2022gdc} for a recent proposal for the case of broken power-laws). 

We now continue with the case $m=4$. The function ${\cal S}_4$ results (see
Appendix \ref{app_Rmmp})
\be
{\cal S}_4\,=\,\sum_{{\cal O}{\cal O}'}
{\cal R}_ {{\cal O}{\cal O}'}^{(2,2)}\,=\, 
 \frac{c_3}{16}\left(\sum_{\cal O \cal O'}\,d_{{\cal O}\,ab}\,d_{{\cal O}'\,cd}\, v_{\perp}^a\,v_{\perp}^b  \, v_{\perp}^c\,v_{\perp}^d\right).
\ee
In this case, this quantity is insensitive to $c_2$: hence it can probe
the function $c_3$ that -- although small -- is the only parameter contributing. 
Proceeding exactly as above, the corresponding  
SNR$_4$ reads
\be
\label{res_snr4}
{\rm SNR}_4 = \left(\sqrt{\frac{T}{1 {\rm year}}}\times \frac{\Omega_{\rm GW}^{(0)}}{{10^{-12}}}\right)\left(\frac{\Sigma_{\mathcal{O}\mathcal{O'}}\,d_{\mathcal{O}\,ab}\,d_{\mathcal{O'}\,cd}\,v_\perp^a\,v_\perp^b\,v_\perp^c\,v_\perp^d}{\frac{4}{5}\,\Sigma_{\mathcal{O}\mathcal{O'}}\,d_{\mathcal{O}\,ab}\,d_{\mathcal{O'}}^{ab}}\right)\left(\int_{f_{\rm min}}^{f_{\rm max}}\,{\rm d}f\,\mathcal{B}_1(f)\right)^{1/2},
\ee

with 
\be
\label{def_BF}
\mathcal{B}_1(f) = 3.9\times 10^{-3}\times c_3^2(f)\,\left(\frac{\Omega_{\rm GW}(f)}{\Omega_{\rm GW}^{(0)}}\right)^2\left(\frac{10^{-12}}{\Sigma_{\rm GW}(f)}\right)^2.
\ee

\begin{figure}[h!]
\centering
  \includegraphics[width = 0.75 \textwidth]{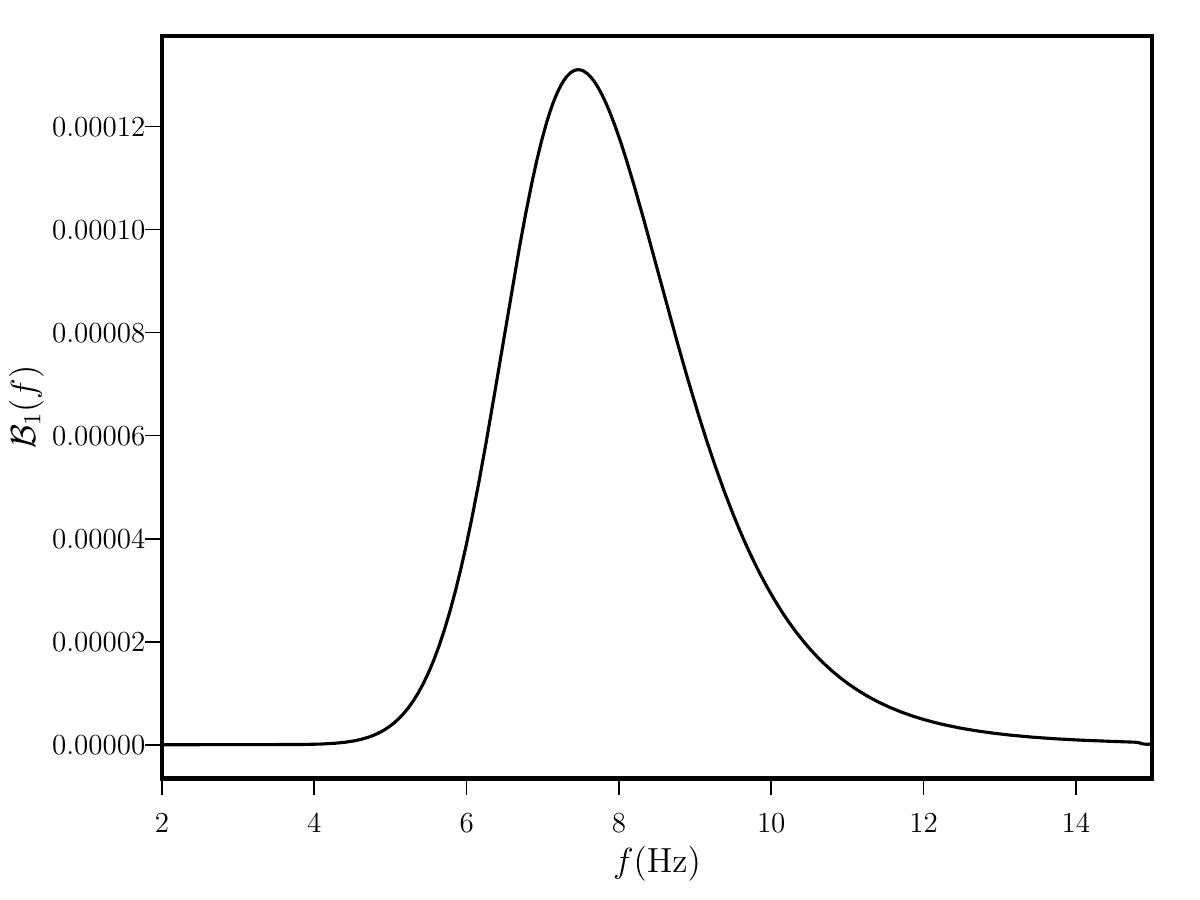}
 \caption{\small{  Plot of the  function ${\cal B}_1(f)$ in eq \eqref{def_BF}, computed for the example 
 of Figures \ref{fig_plot_exam1},  \ref{fig_plot_examc23}. 
   }}
 \label{fig:plotSNRf4}
\end{figure}

 Again, the profile of the function (see Fig. \ref{fig:plotSNRf4})
is peaked at frequencies around the maximal sensitivity of ET, and it is
slightly slanted towards the right, because of the `high-pass' behaviour
of the weighted function $c_3^2$. This time, the integral can be numerically computed to be
\be
\label{res_int}
\left( 
\int_{1\,{\rm Hz}}^{10^4\,{\rm Hz}} \frac{d f}{\rm 1\,Hz}\,
 {\cal B}_1(f)
 \right)^{1/2}\,=\,3.96\times 10^{-4}\,,
\ee 
making it harder to detect the effects of $c_3$ with respect to $c_2$. The remaining cases of $m\neq 2, \,4$ can be treated analogously to these ones, and we do not go through them explicitly.

\smallskip
As a last example, we compute the signal-to-noise ratio ${\rm  SNR}_4$ for 
the case of double broken power-law of eq \eqref{def_inBPL2}, that is, the profile with resonance 
in Fig \ref{fig:plot2DBPL}, as  discussed in Section \ref{sec_reso}. By making use of 
eq \eqref{def_Ois}, 
we convert
the GW intensity of eq \eqref{def_inBPL2} into GW energy density.
 The corresponding profile  
 is
\be
{\Omega}_{\rm GW}(f)\,=\,
{\Omega}_{1}\,
\left(\frac{f}{f_{1}}\right)^{\gamma} \left[
  1+\left( \frac{f}{f_\star}\right)^{\frac{1}{\kappa}}
  \right]^{-\kappa (\gamma+\delta)}+
{\Omega}_{2}  \,\left(\frac{f}{f_{\rm 2}}\right)^{\gamma_1} \left[
  1+\left( \frac{f}{f_3}\right)^{\frac{1}{\kappa_1}}
  \right]^{-\kappa_1 (\gamma_1+\delta_1)}\,,
  \label{def_inBPL2ab}
\ee
where ${\Omega}_{1,2}$ are normalization factors,  and the meaning and implications
of the remaining quantities are discussed after eq \eqref{def_inBPL2}.

\begin{figure}[h!]
\centering
    \includegraphics[width = 0.49 \textwidth]{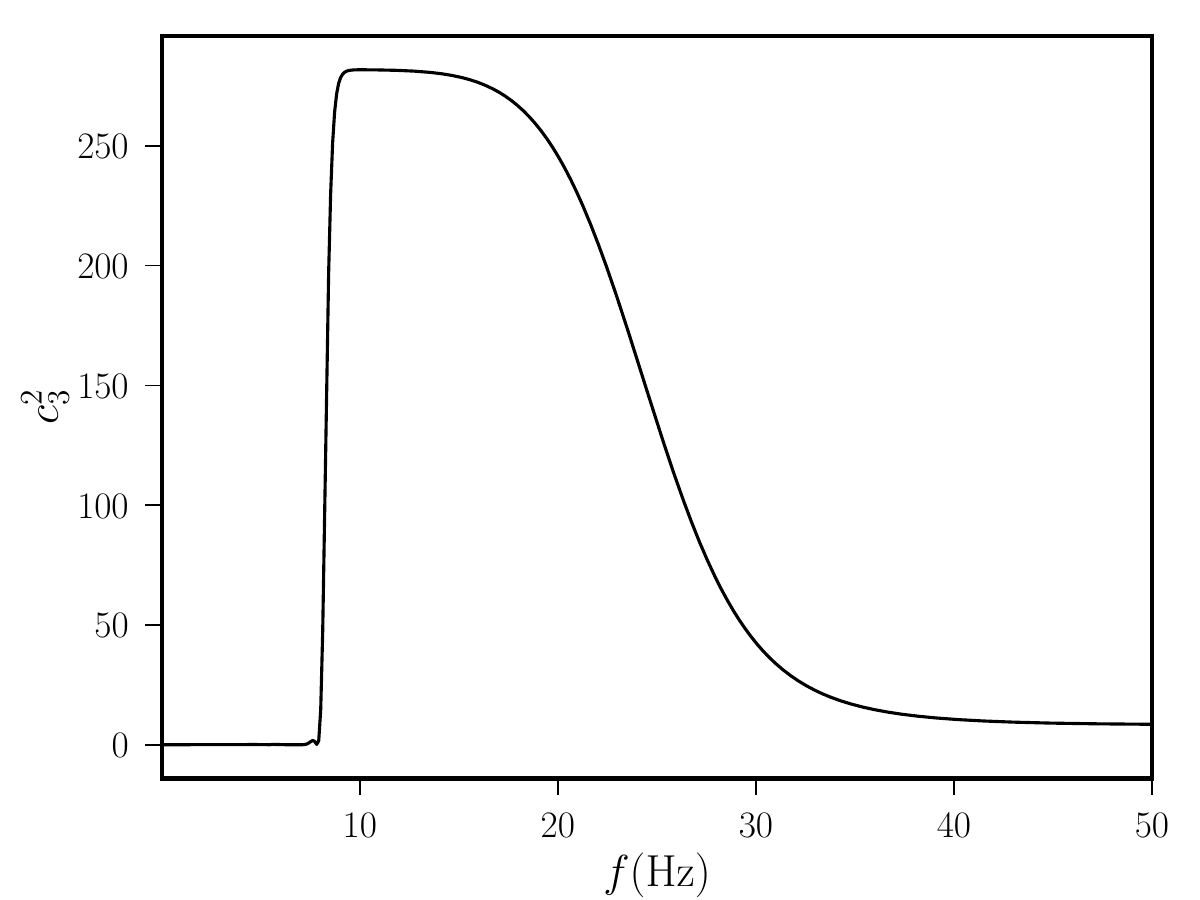} 
        \includegraphics[width = 0.49  \textwidth]{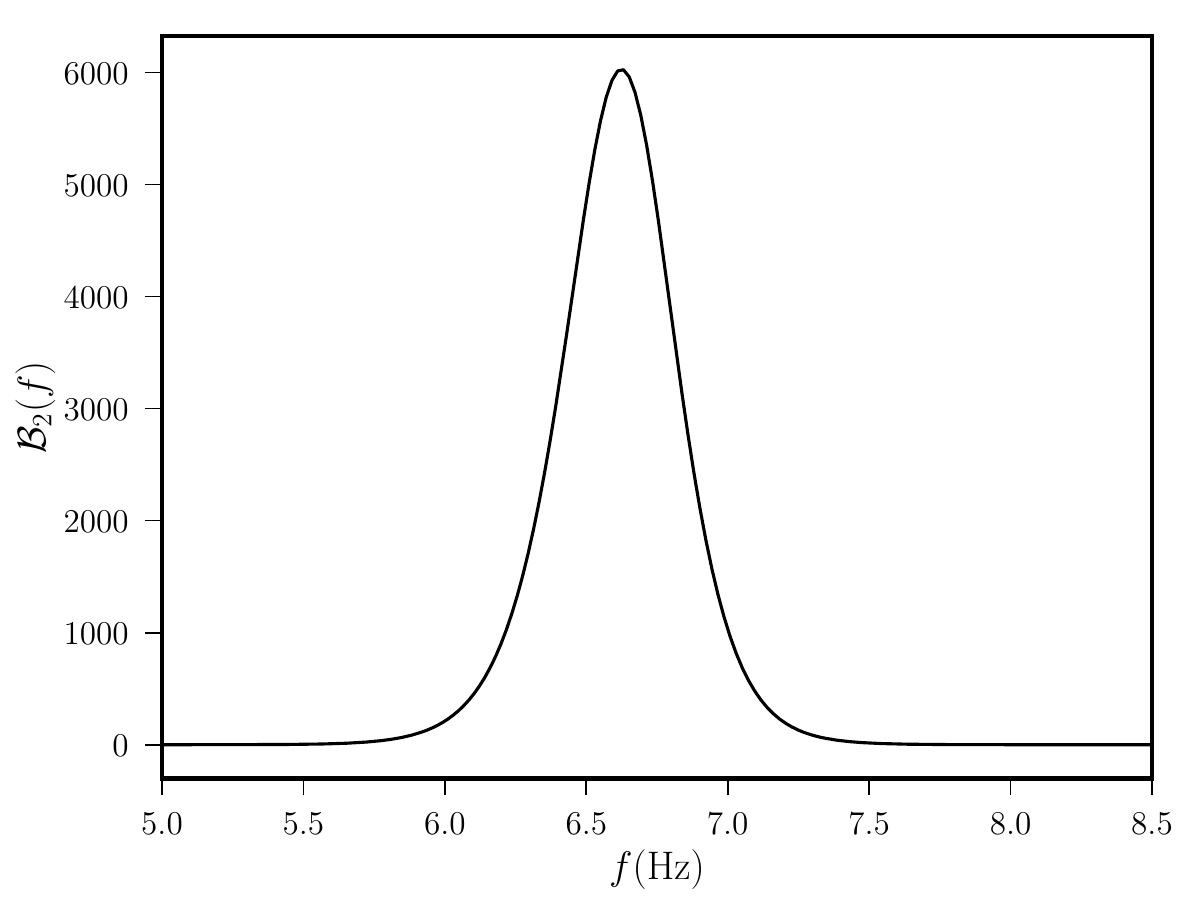}
 \caption{\small{ 
 {\bf Left panel:} The profile of the quantity $c_3^2$ in the scenario with energy profile 
 \eqref{def_inBPL2ab}, 
 represented in Fig \ref{fig:plot2DBPL}.  It is 
 non-vanishing only for a limited range of frequencies. 
  {\bf Right panel:} The function ${\cal B}_2(f)$ of eq \eqref{def_CF}.
   }}
 \label{fig:plotSNRf4a}
\end{figure}

We express the formula for the  SNR for the case $m=4$ as

\be
\label{res_snr4a}
{\rm SNR}_4 = \left(\sqrt{\frac{T}{1 {\rm year}}}\times \frac{\Omega_{\rm GW}^{(0)}}{{10^{-12}}}\right)\left(\frac{\Sigma_{\mathcal{O}\mathcal{O'}}\,d_{\mathcal{O}\,ab}\,d_{\mathcal{O'}\,cd}\,v_\perp^a\,v_\perp^b\,v_\perp^c\,v_\perp^d}{\frac{4}{5}\,\Sigma_{\mathcal{O}\mathcal{O'}}\,d_{\mathcal{O}\,ab}\,d_{\mathcal{O'}}^{ab}}\right)\left(\int_{f_{\rm min}}^{f_{\rm max}}\,{\rm d}f\,\mathcal{B}_2(f)\right)^{1/2},
\ee
with $\Omega_{\rm GW}^{(0)}$ being a `fiducial'  reference scale for the amplitude of the SGWB energy density, and
\be
\label{def_CF}
\mathcal{B}_2(f) = 3.9\times 10^{-3}\times c_3^2(f)\,\left(\frac{\Omega_{\rm GW}(f)}{\Omega_{\rm GW}^{(0)}}\right)^2\left(\frac{10^{-12}}{\Sigma_{\rm GW}(f)}\right)^2.
\ee
The function ${\cal B}_2(f)$
depends on $c_3^2$, which  assumes large values. 
 The profile of ${\cal B}_2(f)$ is  peaked for values   of
 frequency at the maximal sensitivity of ET (see Fig. \ref{fig:plotSNRf4a}).
\textcolor{black}{The frequency  integral  in eq \eqref{res_snr4a} can be evaluated to be
\be
\label{res_int2}
\left( 
\int_{1\,{\rm Hz}}^{10^4\,{\rm Hz}} \frac{d f}{\rm 1\,Hz}\,
 {\cal B}_2(f)
 \right)^{1/2}\,=\,3.14\times 10^3.
\ee}
Apparently, having pronounced resonances or features seems to enhance the
detectability of anisotropic signals, a fact  pointed out recently in \cite{Dimastrogiovanni:2022eir}.

 \section{\color{black}Conclusions} 
 \label{sec_conc}

We studied the response function of the Einstein Telescope to
kinematic anisotropies of the stochastic gravitational wave
background. For the first time we did not  assume a factorizable Ansatz
for the Doppler effects, nor did we
assume the  limit 
of small velocity among frames. We applied
our findings to quantitatively
study the response functions for three
well motivated examples
of gravitational wave
background profiles: power-law, broken
power-law, and
 models with resonances motivated by primordial black hole scenarios. 
We then
 derived the signal- to-noise ratio associated
 with   an optimal estimator for the detection of non-factorizable kinematic anisotropies.
 We analyzed  the signal-to-noise ratio for   some representative examples of broken and 
 doubly broken power-law profiles.
 
 Our work can be extended in several directions. First of all, it would
 be interesting to study further examples of realistic background
 profiles, to investigate more systematically  how Doppler kinematic
 effects depend on the background profile, and which scenarios lead to higher signal-to-noise ratio and are easier to detect.
 Also, it would be useful to quantify (as in  \cite{Mentasti:2020yyd}) corrections to the assumption
 of negligible vertex distance, as discussed towards the end
 of section  \ref{sec_setup}, and investigate whether those corrections can be important
 for specific background profiles.
 
  At the level of applications, the detection and precise
  measurements of kinematic anisotropies can represent
  a new indirect avenue for characterizing the properties
  of the stochastic gravitational wave background, and
  its sources. Suppose in fact that 
   the yet-to-be detected  stochastic  background
   is made of different sources: their possible different
   speeds with respect to us make a difference in their contributions
   to Doppler anisotropies, and a measurement of the latter
    might allow us to distinguish among  sources.
   These and other fascinating questions are left to future work.

 \subsection*{\color{black}Acknowledgments} 

We are
 partially funded by the STFC grant ST/T000813/1.  
 For the purpose of open access, the authors have applied a Creative Commons Attribution (CC BY) licence to any Author Accepted Manuscript version arising.

\newpage
\begin{appendix}

\section{\color{black} Proof of equation \eqref{res_OOp}}
\label{app_A}

The aim of this Appendix is to prove eq \eqref{res_OOp}.
  We write 

\be
 {\cal R}_ { {\cal O}\,{\cal O}'}(f, t, t')
 \,=\,{d^{ab}_{{\cal O}}(t)
 \,
d^{cd}_{{\cal O}'}(t)}
\,
\Gamma_{abcd}(f, \beta, \hat v)\,,
\ee
with
\be
\Gamma_{abcd}(f, \beta, \hat v)\,=\,
\sum_{\lambda} 
\int \frac{d^2 \hat n}{4 \pi}
 {\bf P}(f,\,\hat n)\,{\bf e}_{ab}^{(\lambda)}(\hat n)
{\bf e}_{cd}^{(\lambda)}(\hat n)\,,
\ee
and, as in \cite{Mentasti:2020yyd}, we neglect separation distance among detectors, and ${\bf P}$ is given by eq \eqref{def_bfP}. Since it is contracted with the $d^{ab}$'s, 
 we will only be interested in the contributions $\Gamma_{abcd}$ which are traceless along the first and last two indexes. Moreover, $\Gamma_{abcd}$  is symmetric under the interchanges $a \leftrightarrow b$,  $c \leftrightarrow d$,  $a b \leftrightarrow c d$. We use the 
  identity in Appendix A of \cite{Domcke:2019zls}: 
\bea
\label{res_id1}
\sum_\lambda\,{\bf e}_{ab}^{(\lambda)}(\hat n)
{\bf e}_{cd}^{(\lambda)}(\hat n)
&=& \left( \delta_{ac}-\hat n_a \hat n_c\right)
\left( \delta_{bd}-\hat n_b \hat n_d\right)
+\left( \delta_{ad}-\hat n_a \hat n_d\right)
\left( \delta_{bc}-\hat n_b \hat n_c\right)
\nonumber
\\
&&-\left( \delta_{ab}-\hat n_a \hat n_b\right)
\left( \delta_{cd}-\hat n_c \hat n_d\right)\,.
\eea
 The quantity we are after can be conveniently separated into
  two parts:
 \bea
 \Gamma_{abcd}\,=\,\Gamma_{abcd}^{\rm (iso)}+\Gamma_{abcd}^{\rm (aniso)}\,,
 \eea
 with
 \bea
\Gamma_{abcd}^{\rm (iso)}&=&
\sum_{\lambda} 
\int \frac{d^2 \hat n}{4 \pi}
\,{\bf e}_{ab}^{(\lambda)}(\hat n)
{\bf e}_{cd}^{(\lambda)}(\hat n) \,,
\\
\Gamma_{abcd}^{\rm (aniso)}&=&
\sum_{\lambda} 
\int \frac{d^2 \hat n}{4 \pi}
\left( {\bf P}-1\right)
\,{\bf e}_{ab}^{(\lambda)}(\hat n)
{\bf e}_{cd}^{(\lambda)}(\hat n)\,.
  \eea
 By symmetry considerations, (see e.g. \cite{Allen:1997ad}) the isotropic part can only be proportional to the combinations
 \bea
\Gamma_{abcd}^{\rm (iso)}\,=\,{b_1} \,\left(\delta_{ac}  \delta_{bd}+\delta_{ad}  \delta_{bc} \right)+b_2\,
 \delta_{ab}  \delta_{cd}\,,
 \eea
 where $b_{1,2}$ are  functions of frequency. Using identity \eqref{res_id1}, and the fact that $\int {d^2 \hat n}/{4 \pi}\,=\,1$, we find that
 \bea
 \delta_{cd}\,\delta_{ab}\,
\Gamma_{abcd}^{\rm (iso)}&=&6 \,b_1 \,  +9\,b_2\,
  \,=\,
  0\,,
 \\
 \delta_{bd}\,\delta_{ac}\,
\Gamma_{abcd}^{\rm (iso)}&=&12 \,b_1 \,  +3\,b_2\,
 \,=\, 4\,.
 \eea
Hence $b_1=2/5$,  $b_2=-4/15 $.  The part proportional to $b_2$
will vanish upon contractions with the $d^{ab}_{\cal O}$ quantities.

 The anisotropic part $\Gamma_{abcd}^{\rm (aniso)}$
  can in principle be proportional to the following combinations:

 \bea
\Gamma_{abcd}^{\rm (aniso)}&=& c_1 \,\left(\delta_{ac}  \delta_{bd}+\delta_{ad}  \delta_{bc} \right)+c_2 \left(
  \delta_{ac} \hat v_b \hat v_d
  + \delta_{ad} \hat v_b \hat v_c
  +\delta_{bc} \hat v_a \hat v_d
   +\delta_{bd} \hat v_a \hat v_c
   \right)
+c_3  \hat v_a \hat v_b \hat v_c \hat v_d
\nonumber
\noindent
\\
&+&c_4\,
 \delta_{ab}  \delta_{cd}\nonumber
+c_5 \left( \delta_{ab} \hat v_c \hat v_d
  + \delta_{cd} \hat v_a \hat v_b
  \right)\,.
 \eea
Considerations like the ones above lead to the identities
 \bea
 c_1&=&\frac{K_1}{8}+\frac{3\,K_2}{4}+\frac{K_3}{8}\,,
\\
    c_2&=&\frac{3\,K_1}{8}-\frac{3\,K_2}{4}-\frac{5\,K_3}{8}\,,
   \\
  c_3&=&\frac{3\,K_1}{8}-\frac{15\,K_2}{4}+\frac{35\,K_3}{8}\,,
 \\
 c_4&=&\frac{K_1}{8}-\frac{5\,K_2}{4}+\frac{K_3}{8}\,,
 \\
  c_5&=&-\frac{5\,K_1}{8}+\frac{9\,K_2}{4}-\frac{5\,K_3}{8}\,.
   \eea
 The quantities $K$ are given in eq \eqref{def_intK} of the main text. They
  vanish when $\beta=0$.

 Assembling the results, we find that
 the structure of the response function is
 \bea
  {\cal R}_ {{\cal O}\,{\cal O}'}(f, t, t')
 &=&\frac45\left(1 +\frac{5}{2} c_1 \right)\,{d^{ab}_{{\cal O}}(t)
 \,
d_{{\cal O}'\,ab}(t)}
+4\,c_2\,{d^{ab}_{{\cal O}}(t)
 \,
d_{{\cal O}'\,bc}(t)}\,{\hat v}^c\,{\hat v}_a
\nonumber
\\
&&
+c_3\,\left(\hat v^a \hat v^b\,d_{ab\,{\cal O}}(t)\right)\,\left(\hat v^c \hat v^d\,d_{cd\,{\cal O}'}(t)\right)\,,
 \eea
 corresponding to eq \eqref{res_OOp} of the main text.

\section{\color{black} Exact formulas for the power-law case}
\label{app_B}

The aim of this Appendix is to present the
general formulas for the coefficients $c_i$
for the power-law Ansatz of section \ref{sec_firex}.
They are as follows (denoting $\beta_{\pm}\,=\,1\pm\beta$):

\medskip

\bea
c_1&=&\frac{\text{$\beta_- $}^{\frac{1}{2}-\frac{\alpha }{2}} \text{$\beta_+ $}^{\frac{1}{2}-\frac{\alpha }{2}} \left(\text{$\beta_+ $}^{\alpha }-\text{$\beta_- $}^{\alpha }\right)}{\alpha  (\text{$\beta_+ $}-\text{$\beta_- $})}-1\,,
\\
\nonumber
\\
  c_2&=&
  \frac{3 \alpha ^2 \beta  \text{$\beta_- $}^{\alpha } (\text{$\beta_- $} \text{$\beta_+ $})^{3/2}-3 \alpha ^2 \beta  \text{$\beta_+ $}^{\alpha } (\text{$\beta_- $} \text{$\beta_+ $})^{3/2}-2 \alpha  (\alpha +1) (\alpha +2) \beta ^3 \text{$\beta_+ $} (\text{$\beta_- $} \text{$\beta_+ $})^{\alpha /2}}{6 \alpha
  (\text{$\beta_- $} \text{$\beta_+ $})^{\frac{\alpha }{2}}
    (\alpha +1) (\alpha +2) \beta ^3 \text{$\beta_+ $}}
  \nonumber
  \\
  && +\frac{6 (\beta +1) \text{$\beta_+ $}^{\alpha } \sqrt{\text{$\beta_- $} \text{$\beta_+ $}}+3 \alpha  \beta  \sqrt{\text{$\beta_- $}} \text{$\beta_+ $}^{\alpha +\frac{3}{2}} (\alpha  \text{$\beta_+ $}+\beta -2)-3 \text{$\beta_+ $}^{3/2} \text{$\beta_- $}^{\alpha +\frac{1}{2}} (\alpha  \beta  (\alpha  \text{$\beta_+ $}+\beta +2)+2)}{6 \alpha
  (\text{$\beta_- $} \text{$\beta_+ $})^{\frac{\alpha }{2}}
    (\alpha +1) (\alpha +2) \beta ^3 \text{$\beta_+ $}}
    \nonumber\\
 \\
 c_3&=&-
\frac{2 \alpha  (\alpha +1) (\alpha +2) (\alpha +3) (\alpha +4) \beta ^5 (\text{$\beta_- $} \text{$\beta_+ $})^{\alpha /2}}{10 (\text{$\beta_- $} \text{$\beta_+ $})^{\frac{\alpha }{2}}\alpha  (\alpha +1) (\alpha +2) (\alpha +3) (\alpha +4) \beta ^5}
\nonumber
\\
&&-
\frac{5 (\alpha  \beta  ((\alpha +1) \beta  ((\alpha +2) \beta  ((\alpha +3) \beta +4)+12)+24)+24) \text{$\beta_- $}^{\alpha } \sqrt{\text{$\beta_- $} \text{$\beta_+ $}}
}{10 (\text{$\beta_- $} \text{$\beta_+ $})^{\frac{\alpha }{2}}\alpha  (\alpha +1) (\alpha +2) (\alpha +3) (\alpha +4) \beta ^5}
\nonumber
\\
&&
+5\frac{ 
(\alpha  \beta  ((\alpha +1) \beta  ((\alpha +2) \beta  ((\alpha +3) \beta -4)+12)-24)+24) \text{$\beta_+ $}^{\alpha } \sqrt{\text{$\beta_- $} \text{$\beta_+ $}}
}
{10 (\text{$\beta_- $} \text{$\beta_+ $})^{\frac{\alpha }{2}}\alpha  (\alpha +1) (\alpha +2) (\alpha +3) (\alpha +4) \beta ^5}\,.
\eea

\medskip

\medskip

\noindent
These expressions are valid for any $\alpha$
and $0<\beta<1$. Notice that in some specific limits
 these power-law results formally diverge, and
the limits give  logarithmic contributions, as in the  $\alpha\to-3$
case discussed
in the main text.
\newpage

\section{\color{black} The coefficients ${\cal R}^{(m m')}$}
\label{app_Rmmp}

We report here the explicit formulas for the coefficients
${\cal R}_ {{\cal O}{\cal O}'}^{(m,m')}$ as discussed in section
\ref{sec_disen}. They are as follows:

\bea
{\cal R}_ {{\cal O}{\cal O}'}^{(0,0)}&=&\left(\frac45 +2 c_1\right)
\,d_{{\cal O}\,ab}\,d^{\,\,\,ab}_{{\cal O}'}
+4\,
 c_2\,v_{\parallel}^a  \,d_{{\cal O}\,ab}\,d^{\,\,\,b}_{{\cal O}'\,d}\,
 v_{\parallel}^d
 \nonumber
 \\
 &&
 +\frac{c_3}{4}\,d_{{\cal O}\,ab}\,d_{{\cal O}'\,cd}\, \left( v_{\perp}^a\,v_{\perp}^b+ 2\, v_{\parallel}^a v_{\parallel}^b \right)  \,\left( v_{\perp}^c\,v_{\perp}^d+ 2\, v_{\parallel}^c v_{\parallel}^d \right)\,,
\\
{\cal R}_ {{\cal O}{\cal O}'}^{(1,0)}&=&2\,
 c_2\,v_{\perp}^a  \,d_{{\cal O}\,ab}\,d^{\,\,\,b}_{{\cal O}'\,d}\,
 v_{\parallel}^d+\frac{c_3}{2}\,d_{{\cal O}\,ab}\,d_{{\cal O}'\,cd}\, v_{\perp}^a\,v_{\parallel}^b  \,\left( v_{\perp}^c\,v_{\perp}^d+ 2\, v_{\parallel}^c v_{\parallel}^d \right)
 \\
 {\cal R}_ {{\cal O}{\cal O}'}^{(2,0)}&=&\frac{c_3}{8}\,d_{{\cal O}\,ab}\,d_{{\cal O}'\,cd}\, v_{\perp}^a\,v_{\perp}^b  \,\left( v_{\perp}^c\,v_{\perp}^d+ 2\, v_{\parallel}^c v_{\parallel}^d \right)
 \,,
  \\
 {\cal R}_ {{\cal O}{\cal O}'}^{(1,1)}&=&c_2\,v_{\perp}^a  \,d_{{\cal O}\,ab}\,d^{\,\,\,b}_{{\cal O}'\,d}\,
 v_{\perp}^d+{c_3}\,d_{{\cal O}\,ab}\,d_{{\cal O}'\,cd}\, v_{\perp}^a\,v_{\parallel}^b  \, v_{\perp}^c\,v_{\parallel}^d\,,
  \\
 {\cal R}_ {{\cal O}{\cal O}'}^{(2,1)}&=&\frac{c_3}{4}\,d_{{\cal O}\,ab}\,d_{{\cal O}'\,cd}\, v_{\perp}^a\,v_{\perp}^b  \, v_{\perp}^c\,v_{\parallel}^d\,,
  \\
 {\cal R}_ {{\cal O}{\cal O}'}^{(2,2)}&=&\frac{c_3}{16}\,d_{{\cal O}\,ab}\,d_{{\cal O}'\,cd}\, v_{\perp}^a\,v_{\perp}^b  \, v_{\perp}^c\,v_{\perp}^d\,.
\eea
The quantities ${\cal R}_ {{\cal O}{\cal O}'}^{(m,m')}$  have the property that $ {\cal R}_ {{\cal O}{\cal O}'}^{(m, m')}\,=\,
{\cal R}_ {{\cal O}'{\cal O}}^{(m', m)}
$.

\section{\color{black}Computation of the SNR$_{m}$, and proof of eq \eqref{def_snrDEM}}
\label{app_snr}

The aim of this appendix is to compute the optimal signal-to-noise
ratio
\be
\label{def_snrA}
{\text{SNR}}_m\,=\,\frac{\langle {\cal C}_m \rangle}{\langle {\cal C}^2_m \rangle^{1/2}}
\,,
\ee
for the quantities defined in section \ref{sec_snr}. We proceed by first evaluating
the numerator, then the denominator.

\smallskip
\noindent
{\bf Evaluating $\langle {\cal C}_m \rangle$:} 
 For any $m\neq 0$, the stationary noise does not contribute to
 $\langle {\cal C}_m \rangle$.  Hence, this key quantity is only 
 sensitive to the anisotropy signal! 
     Collecting
 results and definitions in the main text, we find that, for non-vanishing index $m$
\bea
\langle {\cal C}_m 
\rangle &=&\frac{1}{2\,T}\,\sum_{{\cal O} {\cal O}'}
 \int_0^T dt \,e^{-2 \pi\,i  m\,\bar f_e\,t} \int_{-\infty}^\infty d f\,d f' \,\tilde Q_{{\cal O} {\cal O}'}(f)\,
 \,\int_{t-\tau/2}^{t+\tau/2} d t'\,d t''\,e^{2 \pi i \left(f-f'\right) \left(t'-t''\right)}
   {\cal R}_{{\cal O} {\cal O}'}(f', t'',t')\,{\cal I}(f')\,,
\nonumber
\\
&=&\frac{1}{2\,T}\,\sum_{{\cal O} {\cal O}'}
 \int_0^T dt \,e^{-2 \pi i  m\,{\bar f}_e\,t} \int_{-\infty}^\infty d f\,d f' \,\tilde Q_{{\cal O} {\cal O}'}(f)\,
\,{\cal I}(f')
\nonumber
\\
&&\times \sum_{m' m''}\,\int_{t-\tau/2}^{t+\tau/2} d t'\,d t''\,
e^{2 \pi i t' \left(f-f' +m' {\bar f}_e\right)}\,
e^{-2 \pi i t'' \left(f-f' -m'' {\bar f}_e\right)}\,
 {\cal R}_{{\cal O} {\cal O}'}^{(m' m'')}(f')\,.
\eea
To handle the nested integrals, 
we start performing the time integrals along $t', t''$.

 We use  the definition of finite-size  $\delta$-function
\be
\int_{t-\tau/2}^{t+\tau/2} d t'\,e^{2 \pi i t' \left(f'-f'' +m' {\bar f}_e\right)}\,=\,\delta_{\tau} \left(f'-f'' +m' {\bar f}_e\right)\,e^{-2 i \pi \,t\,\left(f'-f''+m' \,{\bar f}_e \right)}\,,
\ee
with $\delta_{\tau} \left(x \right)$ given by
\be
\delta_{\tau} \left(x \right)\,\equiv \,\frac{\sin{(\pi\,   x\,\tau)}}{\pi x}
\hskip1cm,\hskip1cm
\lim_{\tau \to \infty}\,\delta_{\tau} \left(x \right)\,=\,\delta_{D} \left(x \right)\,,
\ee
and $\delta_D$ being the Dirac delta. 
Then we get the expression
\bea
\langle {\cal C}_m 
\rangle &=&
\frac{1}{2\,T}\,\sum_{{\cal O} {\cal O}'}\,\sum_{m' m''}
 \int_0^T dt \,e^{-2 \pi i  m\,{\bar f}_e\,t} 
 \int_{-\infty}^\infty d f\,d f' \,\tilde Q_{{\cal O} {\cal O}'}(f)\,
\,{\cal I}(f')
 {\cal R}_{{\cal O} {\cal O}'}^{(m' m'')}(f')
 \nonumber
\\
&&\times
\delta_{\tau} \left(f-f' +m' {\bar f}_e\right)\,
\delta_{\tau} \left(f-f' -m'' {\bar f}_e\right)\,e^{2 i \pi \,t\,\left(m'+m''\right) \,{\bar f}_e }\,.
\eea
Since 
 ${\bar f}_e$ is
much smaller than the frequency  $f$ of GW,
we can neglect the $m \,{\bar f}_e$ contributions in the argument of the $\delta_{\tau}$ functions. Moreover, $\tau$ is much longer than the inverse of GW frequencies. Hence we can
treat one of the $\delta_\tau$ as Dirac delta-function $\delta_D$, and obtain 
\bea
  \label{snr_num}
\langle {\cal C}_m 
\rangle &=&
\frac{\tau}{2}\,\sum_{{\cal O} {\cal O}'}\,
 \int_{-\infty}^\infty d f\,\tilde Q_{{\cal O} {\cal O}'}(f)\,
 {\cal S}_{{\cal O} {\cal O}'}(f)
 \,{\cal I}(f),
\eea
for the numerator of eq \eqref{def_snrA}, with
\be
{\cal S}^{(m)}_{{\cal O} {\cal O}'}(f)\,=\,
\sum_{m', m''=-2}^2\,\delta_{K}(m-m'-m'')
 {\cal R}_{{\cal O} {\cal O}'}^{(m',\,m'')}(f)\,,
\ee
and $\delta_K$ being the Kronecker delta.

\smallskip
\noindent
{\bf Evaluating $\langle {\cal C}^2_m \rangle^{1/2}$:} 
To evaluate the denominator  of eq \eqref{def_snrA}, we 
work under the hypothesis of noise-domination in eq \eqref{def_2ptlas}, and compute
the variance of the noise. The steps are very similar to the
previous ones, and already carried out in 
 \cite{Mentasti:2020yyd}, section 3.2. We   report the result of the calculation:
 \be
 {\langle {\cal C}^2_m \rangle}\,=\,
 \frac{\tau^2}{4\,T}\,\sum_{{\cal O} {\cal O}'}\,
  \int_{-\infty}^\infty d f\,|\tilde Q_{{\cal O} {\cal O}'}(f)|^2\,
  { N}_{{\cal O} }(f)\,{ N}_{{\cal O}' }(f)\,.
  \label{snr_den}
 \ee
We refer the reader to  \cite{Mentasti:2020yyd} for details.

\smallskip
\noindent
{\bf Estimating the optimal  ${\text{SNR}}_m$.} We now collect the results, and assume
that the detector noise ${ N}_{{\cal O} }\,=\,{ N}$ is the same for all the non-null channels.  
 The expression for the SNR is obtained by combining eqs \eqref{snr_num} and \eqref{snr_den}:

\bea
{\text{SNR}}_m&=&\sqrt{T}\,
\frac{\sum_{{\cal O} {\cal O}'}
 \int_{-\infty}^\infty d f\,\tilde Q_{{\cal O} {\cal O}'}(f)\,
 {\cal S}_{{\cal O} {\cal O}'}(f)\,{\cal I}(f)
 }{\left( 
   \int_{-\infty}^\infty d f\,\sum_{{\cal O} {\cal O}'} |\tilde Q_{{\cal O} {\cal O}'}(f)|^2\,
   N^2  (f)
 \right)^{1/2}}\,,
 \nonumber
 \\
 &=&\sqrt{2 T}\,
\frac{\sum_{{\cal O} {\cal O}'}
 \int_{0}^\infty d f\,\tilde Q_{{\cal O} {\cal O}'}(f)\,
 {\cal S}_{{\cal O} {\cal O}'}(f)\,{\cal I}(f)
 }{\left( 
   \int_{0}^\infty d f\,\sum_{{\cal O} {\cal O}'} |\tilde Q_{{\cal O} {\cal O}'}(f)|^2\,
   N^2  (f)
 \right)^{1/2}}\,,
\label{def_snrB} 
\eea
where in the second line we perform an integration only over positive frequencies (hence the $\sqrt{2}$ factor in front). 
We   determine the optimal value for the filter $\tilde Q_{{\cal O} {\cal O}'}(f)$,
using standard techniques based on Wiener filtering \cite{Maggiore:2007ulw}. We introduce
a positive-definite scalar product $\left[ \dots\right]$, defined as:
\be
\left[ A_{{\cal O} {\cal O}'}(f), \,B_{{\cal O} {\cal O}'}(f)\right]\,\equiv\,\,\sum_{{\cal O} {\cal O}'} \int_0^{\infty} d f\,A^*_{{\cal O} {\cal O}'}(f) \,B_{{\cal O} {\cal O}'}(f)\, N^2  (f)\,.
\ee
Using this scalar product, we re-express \eqref{def_snrB} as

\be
\label{def_snrC}
{\text{SNR}}_m\,=\,\sqrt{2 T}\,
\frac{ \left[ \tilde Q_{{\cal O} {\cal O}'}(f), \, {\cal S}_{{\cal O} {\cal O}'}(f)\,{\cal I}(f)/N^2(f)\right]
 }{ 
   \left[\tilde Q_{{\cal O} {\cal O}'}(f),  \tilde Q_{{\cal O} {\cal O}'}(f) \right]^{1/2}}\,.
\ee
This quantity is maximised choosing a filter
$\tilde Q_{{\cal O} {\cal O}'}(f)\,=\,  {\cal S}_{{\cal O} {\cal O}'}(f)\,{\cal I}(f)/N^2(f)$. Using it, the optimal signal-to-noise ratio
 results
 \be
 \label{def_snrD}
{\text{SNR}}_m\,=\,\sqrt{2T}\,
\left( 
\int_{0}^\infty d f\,
\left| \sum_{{\cal O} {\cal O}'}\, {\cal S}^{(m)}_{{\cal O} {\cal O}'}(f)\,\,\frac{{\cal I} (f)}{N(f)}\right|^2
 \right)^{1/2}\,,
 \ee
hence demonstrating eq \eqref{def_snrDEM}.

\end{appendix}

{\small


\providecommand{\href}[2]{#2}\begingroup\raggedright\endgroup

}

\end{document}